\let\KernelLabel\label
\documentclass[aip,pof,reprint,amsmath,amssymb]{revtex4-2}
\let\label\KernelLabel
\makeatletter
\let\ltx@label\label
\makeatother

\usepackage[T1]{fontenc}
\usepackage[utf8]{inputenc}
\usepackage{bm,mathtools}
\usepackage{graphicx}
\usepackage{booktabs}
\usepackage[colorlinks=true,citecolor=blue,urlcolor=blue,linkcolor=black]{hyperref}
\newcommand{\dd}{\mathrm{d}}
\newcommand{\e}{\mathrm{e}}
\newcommand{\Aacc}{\mathcal{A}}
\newcommand{\Zens}{\mathcal{Z}}
\renewcommand{\arraystretch}{1.08}
\begin{document}

\title{A Zero-Flux Material Cylinder in the Unsteady Kiknadze--Krasnov Vortex}
\author{Khalid M. Saqr}
\email{k.saqr@aast.edu}
\affiliation{Mechanical Engineering Department, College of Engineering and Technology, Arab Academy for Science, Technology and Maritime Transport, Alexandria 1029, Egypt}
\thanks{ORCID 0000-0002-3058-2705}

\begin{abstract}
In viscous flow, the circulation enclosed by a deterministic material loop generally changes through vorticity diffusion. It is shown that the source-bearing unsteady Kiknadze--Krasnov vortex possesses a distinguished cylindrical material surface on which this viscous circulation transfer vanishes identically. For a single distributed mode, the scaled squared radius $x=\beta(t)r^2$ satisfies
\[
\frac{\mathrm D x}{\mathrm D t}
=
4\nu\beta(t)\big[(s-1)-x\big],
\]
where $\beta(t)$ is the inverse squared radial scale, $\nu$ is the kinematic viscosity, and $s$ is the incomplete-gamma shape parameter. For $s>1$, the cylinder $x=s-1$ is transported exactly by the radial flow and preserves the inside/outside ordering of fluid trajectories. The same factor $(s-1)-x$ governs the radial vorticity gradient and the material circulation-transfer rate. Consequently, the magnitude of the distributed axial vorticity $|\omega_z|$ attains its annular maximum on this material cylinder, $\partial_r\omega_z=0$ there, and the enclosed circulation remains exactly $\Gamma P(s,s-1)$ for every admissible prescribed strain history. Neighbouring material cylinders continue to exchange circulation viscously, while the distinguished moving cylinder carries a fixed fraction of the distributed circulation throughout the admissible evolution. In the steady source--strain limit, the surface reduces to the radial stagnation cylinder. The complete velocity and pressure fields, exact radial-scale evolution, admissible parameter ranges, and qualifications for multiple radial scales are also derived. The resulting solution-specific material-circulation invariant provides an exact benchmark for numerical material-surface tracking, vortex-core evolution, and viscous circulation transport, with the principal analytical results additionally supported by machine-checked formal verification.
\end{abstract}

\maketitle

\section{Introduction}\label{sec:intro}

In exact viscous vortex solutions, the coupling among advection, vortex stretching and diffusion is evaluated analytically, without spatial discretisation or closure modelling. Viscous spreading is described by the Lamb--Oseen vortex, a stationary strain--diffusion balance is represented by the Burgers vortex, and arbitrary initial vorticity in an irrotational strain field is accommodated by Kambe's axisymmetric construction \cite{Oseen1911,Burgers1948,Kambe1984}. Exact strained vortices and their relation to turbulence modelling have been reviewed by Ohkitani \cite{Ohkitani2004}.

Unsteady axisymmetric viscous vortices with radial transport also have an established exact-solution lineage. Bellamy-Knights constructed unsteady two-cell viscous vortices in which radial flux is an explicit solution parameter \cite{BellamyKnights1970}. Craik subsequently derived exact axisymmetric Navier--Stokes vortices subject to strain together with suction or injection, exhibiting explicitly the competing roles of diffusion, advection and vortex stretching \cite{Craik2009}. These results provide important precedents for exact vortices with radial transport, while the Kiknadze--Krasnov family considered below has the particular incomplete-gamma circulation structure used in the present material-surface analysis.

A broader axisymmetric solution was reported by Kiknadze and Krasnov \cite{KiknadzeKrasnov1986}. Time-dependent irrotational strain is combined with radial line-source or line-sink motion and azimuthal circulation. The distributed swirl is expressed by incomplete gamma functions. In the notation adopted here,
\begin{equation}
 u_r=-\frac{a(t)}{2}r+\frac{q}{r},
 \qquad
 u_z=a(t)z+b(t),
 \label{eq:intro_meridional}
\end{equation}
and a finite-circulation distributed component is
\begin{equation}
 u_\theta=\frac{\Gamma}{2\pi r}P\!\left(s,\beta(t)r^2\right),
 \qquad
 s=1+\frac{q}{2\nu}.
 \label{eq:intro_swirl}
\end{equation}
Here $t$ is time; $r$ and $z$ are the radial and axial coordinates; $u_r$, $u_z$ and $u_\theta$ are the radial, axial and azimuthal velocity components; $a(t)$ is the prescribed axial strain rate; $q$ is the constant line-source or line-sink strength; and $b(t)$ is a uniform axial-velocity component. The quantity $\Gamma$ is the total distributed circulation, $\beta(t)>0$ is its inverse squared radial scale, $\nu$ is the kinematic viscosity, and $P(s,x)=\gamma(s,x)/\Gamma(s)$ is the regularised lower incomplete gamma function with shape parameter $s$; $\gamma(s,x)$ and $\Gamma(s)$ denote the lower incomplete and complete gamma functions.

An independent central $1/r$ circulation and finite superpositions of distributed components are also admitted. Restricted forms have been used in subsequent applications \cite{Kiknadze2009,Bockeria2013,Beglov2025}. The pressure field, compatibility conditions, admissible domains and parameter ranges are derived here from the governing equations for the combined strain--source--swirl field.

Viscous circulation transport is examined within this exact solution. For an ordinary deterministic material loop, viscosity generally modifies circulation through vorticity diffusion. Stochastic Lagrangian formulations provide a different circulation-conservation statement for Navier--Stokes dynamics. Constantin and Iyer obtained a stochastic representation of incompressible Navier--Stokes dynamics \cite{ConstantinIyer2008}, and Eyink formulated the associated stochastic Kelvin theorem in variational form \cite{Eyink2010}. Those results concern circulation transported through stochastic Lagrangian trajectories or their averages. The present solution-specific analysis concerns an ordinary deterministic material loop in an exact viscous vortex whose motion with the fluid is compatible with identically vanishing viscous circulation transfer.

General material barriers to diffusive transport have been characterized through transport functionals and variational conditions \cite{HallerKarraschKogelbauer2018}. Vortex boundaries have likewise been formulated as material barriers to diffusive vorticity transport in two-dimensional flows \cite{KatsanoulisFarazmandSerraHaller2020}, and objective transport-barrier theory has been extended to dynamically active vector fields, including vorticity, in three dimensions \cite{HallerKatsanoulisHolznerFrohnapfelGatti2020}. The present analysis identifies a material surface in the source-bearing Kiknadze--Krasnov solution with exactly zero deterministic viscous circulation transfer for every admissible prescribed strain history. Classification under the variational barrier definitions requires a separate evaluation of the corresponding transport functional.

For a single distributed KK component, the scaled squared radius
\begin{equation}
 x=\beta(t)r^2
 \label{eq:intro_x}
\end{equation}
is governed along a fluid trajectory by
\begin{equation}
 \frac{\mathrm D x}{\mathrm D t}
 =4\nu\beta(t)\big[(s-1)-x\big].
 \label{eq:intro_Dx}
\end{equation}
Here $\mathrm D/\mathrm Dt$ denotes the material derivative following the trajectory. In Eq.~\eqref{eq:intro_Dx}, the factor $(s-1)-x$ controls scaled radial material motion and also appears in the radial vorticity gradient and the material circulation-transfer rate. For $s>1$, the cylinder $x=s-1$ is transported by the fluid, carries the maximum magnitude of the distributed vorticity, and has identically zero viscous circulation transfer throughout the admissible evolution. It therefore encloses the time-independent fraction $P(s,s-1)$ of the distributed circulation for arbitrary admissible time-dependent strain.

In the steady source--strain state, the cylinder reduces to the surface where radial stagnation coincides with the vorticity-magnitude maximum, identified by Rajamanickam and Weiss \cite{RajamanickamWeiss2021}. The result established here is the unsteady persistence of the coupled material-motion and zero-transfer conditions when the physical cylinder moves according to the history-dependent scale $\beta(t)$. To the author's knowledge, this persistent unsteady coincidence has not previously been identified for the Kiknadze--Krasnov family.

The surface is termed an exact zero-flux material circulation surface. This terminology refers specifically to the deterministic material-motion identity and the vanishing viscous circulation rate derived below. Variational active-transport-barrier classification requires a separate assessment of the corresponding transport functional on neighbouring surfaces.

For distributed-mode index $i$, the inverse squared scale $\beta_i(t)$ satisfies
\begin{equation}
 \dot\beta_i-a(t)\beta_i+4\nu\beta_i^2=0.
 \label{eq:intro_beta}
\end{equation}
An overdot denotes differentiation with respect to time.
Finite-time concentration is classified from the explicit scale solution. The established singular Burgers-vortex behaviour is recovered under reciprocal-time strain \cite{Moffatt2000,MaekawaMiuraPrange2019}. For non-zero linear background strain, the whole-space kinetic energy is infinite; the concentration analysis is therefore restricted to the KK class.

The admissible parameter ranges are determined from axis regularity, finite distributed circulation and convergence of the enstrophy integral. The velocity and pressure fields are verified by substitution into the mass and momentum balances.
\section{Coordinate system, reference frame and governing equations}\label{sec:governing}

\subsection{Physical setting and notation}
A homogeneous Newtonian fluid is considered, with constant density $\rho>0$, dynamic viscosity $\mu>0$, and kinematic viscosity
\begin{equation}
 \nu=\frac{\mu}{\rho}>0.
\end{equation}
The motion is described relative to a fixed right-handed inertial reference frame. Cylindrical coordinates $(r,\theta,z)$ are introduced according to $X=r\cos\theta$ and $Y=r\sin\theta$, with basis vectors $\bm e_r$, $\bm e_\theta$ and $\bm e_z$. The $z$ axis is the symmetry and vortex axis. The velocity field is
\begin{equation}
 \bm u=u_r\bm e_r+u_\theta\bm e_\theta+u_z\bm e_z.
\end{equation}
Axisymmetry is imposed through $\partial/\partial\theta=0$, with azimuthal motion permitted. The pressure is $p=p(r,z,t)$. Non-conservative body forces are set to zero; any conservative body-force potential can be absorbed into a modified pressure.

Positive $a(t)$ denotes axial stretching with associated radial convergence. The corresponding KK coefficient $C_0(t)$ is
\begin{equation}
 C_0(t)=-\frac{a(t)}{2}.
 \label{eq:C0map}
\end{equation}
\subsection{Integral conservation of mass}
An arbitrary fixed control volume $V$ is considered, with boundary $\partial V$ and outward normal $\bm n$. The symbols $\dd V$ and $\dd A$ denote differential volume and surface-area elements. Conservation of mass is written
\begin{equation}
 \frac{\dd}{\dd t}\int_V\rho\,\dd V
 +\int_{\partial V}\rho\bm u\cdot\bm n\,\dd A=0.
\end{equation}
For constant density, the divergence theorem gives
\begin{equation}
 \nabla\cdot\bm u=0.
 \label{eq:continuity_vector}
\end{equation}

\subsection{Integral conservation of momentum}
Linear momentum conservation is
\begin{equation}
 \frac{\dd}{\dd t}\int_V\rho\bm u\,\dd V
 +\int_{\partial V}\rho\bm u(\bm u\cdot\bm n)\,\dd A
 =\int_{\partial V}\bm\sigma\bm n\,\dd A,
\end{equation}
where the Cauchy stress tensor of an incompressible Newtonian fluid is
\begin{equation}
 \bm\sigma=-p\bm I+2\mu\bm D,
 \qquad
 \bm D=\frac12\left[\nabla\bm u+(\nabla\bm u)^{\mathsf T}\right].
\end{equation}
Here $\bm I$ is the identity tensor and $\bm D$ is the rate-of-strain tensor.
Application of mass conservation and the divergence theorem gives
\begin{equation}
 \frac{\partial\bm u}{\partial t}
 +(\bm u\cdot\nabla)\bm u
 =-\frac{1}{\rho}\nabla p+\nu\nabla^2\bm u.
 \label{eq:NS_vector}
\end{equation}
The full unsteady inertial and viscous terms are retained, including the azimuthal contributions to momentum transport.

\subsection{Axisymmetric equations}
Under $\partial/\partial\theta=0$, continuity becomes
\begin{equation}
 \frac1r\frac{\partial(ru_r)}{\partial r}
 +\frac{\partial u_z}{\partial z}=0.
 \label{eq:continuity_cyl}
\end{equation}
In the component equations below, subscripts following a comma denote partial differentiation; for example, $u_{r,rr}=\partial^2u_r/\partial r^2$. The radial momentum equation is
\begin{align}
 \frac{\partial u_r}{\partial t}
 +u_r\frac{\partial u_r}{\partial r}
 +u_z\frac{\partial u_r}{\partial z}
 -\frac{u_\theta^2}{r}
 =&-\frac1\rho\frac{\partial p}{\partial r}\nonumber\\
 &+\nu\left(
 u_{r,rr}+\frac1r u_{r,r}-\frac{u_r}{r^2}+u_{r,zz}
 \right).
 \label{eq:radial_NS}
\end{align}
The azimuthal equation is
\begin{align}
 \frac{\partial u_\theta}{\partial t}
 +u_r\frac{\partial u_\theta}{\partial r}
 +u_z\frac{\partial u_\theta}{\partial z}
 +\frac{u_ru_\theta}{r}
 =\nu\left(
 u_{\theta,rr}+\frac1r u_{\theta,r}-\frac{u_\theta}{r^2}+u_{\theta,zz}
 \right),
 \label{eq:azimuthal_NS}
\end{align}
and the axial equation is
\begin{align}
 \frac{\partial u_z}{\partial t}
 +u_r\frac{\partial u_z}{\partial r}
 +u_z\frac{\partial u_z}{\partial z}
 =-\frac1\rho\frac{\partial p}{\partial z}
 +\nu\left(u_{z,rr}+\frac1r u_{z,r}+u_{z,zz}\right).
 \label{eq:axial_NS}
\end{align}

\subsection{Natural domains}
Two domains are distinguished. The regular-axis domain is
\begin{equation}
 \Omega_0=\{(r,\theta,z):r\ge0,\ 0\le\theta<2\pi,\ z\in\mathbb R\},
\end{equation}
where a smooth axisymmetric velocity satisfies $u_r(0,z,t)=u_\theta(0,z,t)=0$ and bounded vorticity and pressure. When a $1/r$ line-source, line-sink or line-circulation term is present, the natural domain is punctured,
\begin{equation}
 \Omega_q=\{(r,\theta,z):r>0,\ 0\le\theta<2\pi,\ z\in\mathbb R\}.
\end{equation}
The axis may alternatively be replaced by a finite inner cylinder $r=R_i>0$.

\begin{figure*}[t]
 \centering
 \includegraphics[width=0.98\textwidth]{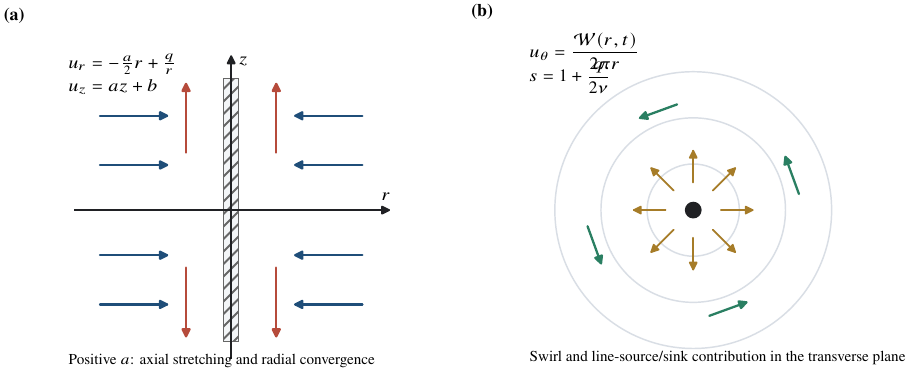}
 \caption{Coordinate system and physical structure of the KK flow. (a) Positive $a(t)$ gives axial stretching and radial convergence; $q/r$ adds line-source or line-sink transport. (b) Transverse-plane view of distributed swirl and the radial source/sink contribution; $\mathcal W(r,t)=2\pi r u_\theta$ is the circulation function and $s=1+q/(2\nu)$.}
 \label{fig:coordinates}
\end{figure*}

The geometry and sign conventions are shown in Fig.~\ref{fig:coordinates}. A line singularity excluded from $\Omega_q$ at the initial time is distinct from a finite-time singularity generated during the evolution.

For non-zero KK strain, the kinetic-energy integral over $\mathbb R^3$ diverges because the velocity grows linearly with spatial distance. Finite-time regularity statements are restricted to the natural KK domain and, where required, to bounded spatial subsets.
\section{Construction of the Kiknadze--Krasnov velocity field}\label{sec:construction}

\subsection{Meridional flow from incompressibility}
An axisymmetric meridional potential flow is prescribed for the KK class,
\begin{equation}
 u_r=u_r(r,t),\qquad u_z=u_z(z,t),\qquad u_\theta=u_\theta(r,t).
 \label{eq:KK_dependencies}
\end{equation}
Equation~\eqref{eq:continuity_cyl} then contains one function of $r,t$ and one of $z,t$. The axial gradient is identified with the prescribed strain rate,
\begin{equation}
 \frac{\partial u_z}{\partial z}=a(t).
\end{equation}
Integration gives
\begin{equation}
 u_z=a(t)z+b(t).
 \label{eq:uz_KK}
\end{equation}
Continuity requires
\begin{equation}
 \frac1r\frac{\partial(ru_r)}{\partial r}=-a(t),
\end{equation}
so
\begin{equation}
 ru_r=-\frac{a(t)}2r^2+q(t),
\end{equation}
and therefore
\begin{equation}
 u_r=-\frac{a(t)}2r+\frac{q(t)}r.
 \label{eq:ur_provisional}
\end{equation}
Incompressibility alone permits $q(t)$. Constancy of $q$ is imposed by compatibility with the fixed swirl profile.

The flux carried by the $q/r$ contribution through a cylindrical surface per unit axial length is
\begin{equation}
 Q_r=\int_0^{2\pi}\frac{q}{r}r\,\dd\theta=2\pi q,
\end{equation}
so $q>0$ denotes an outward line source and $q<0$ an inward line sink.

\subsection{Angular-momentum equation}
The specific axial angular momentum is defined as
\begin{equation}
 L(r,t)=r u_\theta(r,t).
 \label{eq:Ldef}
\end{equation}
Because the KK swirl is independent of $z$, Eq.~\eqref{eq:azimuthal_NS} becomes
\begin{equation}
 \frac{\partial L}{\partial t}
 +u_r\frac{\partial L}{\partial r}
 =\nu\left(\frac{\partial^2L}{\partial r^2}-\frac1r\frac{\partial L}{\partial r}\right).
 \label{eq:Ltransport}
\end{equation}
Once $u_r$ is prescribed, this is a linear advection--diffusion equation for angular momentum.

The radial--temporal variable is introduced as
\begin{equation}
 x=\beta(t)r^2,
 \label{eq:xdef}
\end{equation}
and a fixed profile $L=F(x)$ is assumed, where $F$ is the angular-momentum profile and a prime denotes differentiation with respect to $x$. The required derivatives are
\begin{equation}
 L_t=\frac{\dot\beta}{\beta}xF',\qquad
 L_r=2\beta rF',\qquad
 L_{rr}-\frac1rL_r=4\beta xF''.
\end{equation}
Substitution of Eq.~\eqref{eq:ur_provisional} gives
\begin{equation}
 \left(\frac{\dot\beta}{\beta}-a\right)xF'
 +2\beta q(t)F'=4\nu\beta xF''.
 \label{eq:preseparation}
\end{equation}
The temporal and radial dependence separate when
\begin{equation}
 \dot\beta-a(t)\beta+4\nu\beta^2=0,
 \label{eq:beta_evolution}
\end{equation}
leaving
\begin{equation}
 2\nu xF''+[2\nu x-q(t)]F'=0.
\end{equation}
For a fixed non-trivial radial profile, compatibility at all times requires
\begin{equation}
 q(t)=q=\text{constant}.
 \label{eq:qconstant}
\end{equation}
The dimensionless source parameter and gamma-function shape parameter are defined by
\begin{equation}
 \chi=\frac{q}{2\nu},\qquad s=1+\chi=1+\frac{q}{2\nu}.
 \label{eq:sdef}
\end{equation}
The profile equation becomes
\begin{equation}
 xF''+(x-\chi)F'=0.
 \label{eq:Fode}
\end{equation}

\subsection{Incomplete-gamma swirl}
With the profile derivative $G(x)=F'(x)$, Equation~\eqref{eq:Fode} gives
\begin{equation}
 \frac{G'}G=\frac{\chi}{x}-1,
\end{equation}
so
\begin{equation}
 G=Kx^\chi\e^{-x}.
\end{equation}
Here $K$ is an integration constant. Integration yields the upper-incomplete-gamma representation
\begin{equation}
 F(x)=K_0+K_1\Gamma(s,x),
 \qquad
 \Gamma(s,x)=\int_x^\infty \xi^{s-1}\e^{-\xi}\,\dd\xi,
 \label{eq:F_upper_gamma}
\end{equation}
where $K_0$ and $K_1$ are integration constants and $\xi$ is a dummy integration variable. This is the characteristic KK swirl structure \cite{KiknadzeKrasnov1986}. The complete and lower incomplete gamma functions are
\begin{equation}
 \Gamma(s)=\int_0^\infty \xi^{s-1}\e^{-\xi}\,\dd\xi,
 \qquad
 \gamma(s,x)=\int_0^x \xi^{s-1}\e^{-\xi}\,\dd\xi.
\end{equation}
For $s>0$, the regularised form is
\begin{equation}
 P(s,x)=\frac{\gamma(s,x)}{\Gamma(s)}
 =1-\frac{\Gamma(s,x)}{\Gamma(s)}.
 \label{eq:Pdef}
\end{equation}
A finite-circulation distributed mode can then be written
\begin{equation}
 u_{\theta,i}(r,t)=\frac{\Gamma_i}{2\pi r}P\!\left(s,\beta_i(t)r^2\right).
 \label{eq:single_distributed_swirl}
\end{equation}
Here $\Gamma_i$ is the total distributed circulation of mode $i$. An independent central line-circulation term $\Gamma_\ell/(2\pi r)$ may be added, where $\Gamma_\ell$ is the line circulation.

Because Eq.~\eqref{eq:Ltransport} is linear, any finite sum of solutions with the same $s$ and imposed $a(t)$ remains exact. The multiscale form is therefore
\begin{equation}
 u_\theta(r,t)=\frac{1}{2\pi r}
 \left[
 \Gamma_\ell+\sum_{i=1}^N\Gamma_i P\!\left(s,\beta_i(t)r^2\right)
 \right],
 \label{eq:multimode_swirl}
\end{equation}
where $N$ is the number of distributed modes, $i=1,\ldots,N$ labels each mode, and every $\beta_i$ satisfies Eq.~\eqref{eq:beta_evolution}. The angular-momentum equation is satisfied exactly by the finite superposition because it is linear for the prescribed meridional flow.

The total far-field circulation is
\begin{equation}
 \Gamma_\infty=\Gamma_\ell+\sum_{i=1}^N\Gamma_i.
 \label{eq:Gamma_inf}
\end{equation}

\subsection{Exact radial-scale evolution}
The squared scale variable $h_i$ is defined by
\begin{equation}
 h_i=\beta_i^{-1}.
\end{equation}
Equation~\eqref{eq:beta_evolution} becomes
\begin{equation}
 \dot h_i+a(t)h_i=4\nu.
 \label{eq:hlinear}
\end{equation}
The accumulated strain is defined by
\begin{equation}
 \Aacc(t)=\int_0^t a(\sigma)\,\dd\sigma,
 \label{eq:Aacc}
\end{equation}
where $\sigma$ is a dummy time variable. Its exact solution is
\begin{equation}
 \frac1{\beta_i(t)}
 =\e^{-\Aacc(t)}
 \left[
 \frac1{\beta_{i0}}+4\nu\int_0^t\e^{\Aacc(\sigma)}\,\dd\sigma
 \right],
 \label{eq:beta_inverse_exact}
\end{equation}
where $\beta_{i0}=\beta_i(0)$ is the initial inverse squared scale. Equivalently,
\begin{equation}
 \beta_i(t)=\frac{\e^{\Aacc(t)}}
 {\beta_{i0}^{-1}+4\nu\displaystyle\int_0^t\e^{\Aacc(\sigma)}\,\dd\sigma}.
 \label{eq:beta_exact}
\end{equation}
For two modes indexed by $i$ and $j$,
\begin{equation}
 \beta_i^{-1}(t)-\beta_j^{-1}(t)
 =\e^{-\Aacc(t)}
 \left[\beta_{i0}^{-1}-\beta_{j0}^{-1}\right].
 \label{eq:scale_separation}
\end{equation}
The ordering of inverse squared scales is preserved. Their separation is governed by the accumulated strain.

The complete velocity field is given by Eqs.~\eqref{eq:uz_KK}, \eqref{eq:ur_provisional}, and \eqref{eq:multimode_swirl}, with constant $q$ and $\beta_i$ from Eq.~\eqref{eq:beta_exact}.
\section{Pressure field, initial conditions, boundaries and admissible flow classes}\label{sec:pressure}

\subsection{Radial and axial pressure gradients}
For
\begin{equation}
 u_r=-\frac{a}{2}r+\frac qr,
\end{equation}
the time derivative and radial derivative are
\begin{equation}
 u_{r,t}=-\frac{\dot a}{2}r,
 \qquad
 u_{r,r}=-\frac a2-\frac{q}{r^2},
\end{equation}
and the radial convective acceleration is
\begin{equation}
 u_ru_{r,r}=\frac{a^2}{4}r-\frac{q^2}{r^3}.
\end{equation}
The cross terms proportional to $aq/r$ cancel. The radial viscous operator also vanishes identically,
\begin{equation}
 u_{r,rr}+\frac1r u_{r,r}-\frac{u_r}{r^2}=0.
\end{equation}
Equation~\eqref{eq:radial_NS} therefore gives
\begin{equation}
 \frac1\rho p_r
 =\left(\frac{\dot a}{2}-\frac{a^2}{4}\right)r
 +\frac{q^2}{r^3}+\frac{u_\theta^2}{r}.
 \label{eq:pr}
\end{equation}
For $u_z=az+b$,
\begin{equation}
 u_{z,t}+u_zu_{z,z}=(\dot a+a^2)z+(\dot b+ab),
\end{equation}
and all axial viscous terms vanish. Hence
\begin{equation}
 \frac1\rho p_z=-(\dot a+a^2)z-(\dot b+ab).
 \label{eq:pz}
\end{equation}

\subsection{Exact pressure field}
Integration gives
\begin{align}
 p(r,z,t)=&\ p_0(t)
 +\rho\left(\frac{\dot a}{4}-\frac{a^2}{8}\right)r^2
 -\frac{\rho q^2}{2r^2}\nonumber\\
 &-\frac\rho2(\dot a+a^2)z^2
 -\rho(\dot b+ab)z
 +\rho\Pi(r,t),
 \label{eq:pressure_total}
\end{align}
where
\begin{equation}
 \Pi(r,t)=\int_{r_{\mathrm{ref}}}^{r}\frac{u_\theta^2(\xi,t)}{\xi}\,\dd\xi
 =\frac1{4\pi^2}\int_{r_{\mathrm{ref}}}^{r}\frac{\mathcal W^2(\xi,t)}{\xi^3}\,\dd\xi,
 \label{eq:swirl_pressure_integral}
\end{equation}
Here $p_0(t)$ is an arbitrary pressure reference, $\Pi(r,t)$ is the swirl-induced pressure potential, $r_{\mathrm{ref}}$ is its lower radial reference, $\xi$ is a dummy radial integration variable, and $\mathcal W(r,t)=2\pi r u_\theta$ is the circulation function. For a regular-axis flow, $r_{\mathrm{ref}}=0$ may be used; a punctured-domain flow requires $r_{\mathrm{ref}}>0$. The swirl-induced pressure is represented by this integral for any finite number of distributed modes, including their cross terms in $u_\theta^2$.

At fixed time and axial coordinate, the near-axis pressure satisfies
\begin{equation}
 \lim_{r\to0^+}r^2p(r,z,t)
 =-\frac{\rho}{2}
 \left(
 q^2+\frac{\Gamma_\ell^2}{4\pi^2}
 \right).
\end{equation}
Thus, on the zero-line-circulation branch $\Gamma_\ell=0$, the leading singular coefficient reduces to $-\rho q^2/2$, so for $q\ne0$, $p\sim-\rho q^2/(2r^2)$. A nonzero central line circulation contributes the additional leading swirl-pressure term $-\rho\Gamma_\ell^2/(8\pi^2r^2)$. These line singularities are prescribed by the constant coefficients $q$ and $\Gamma_\ell$ and are present in the initial field.

\subsection{Regular-axis branch}
Inclusion of the symmetry axis requires $u_r(0)=u_\theta(0)=0$, which implies
\begin{equation}
 q=0,\qquad \Gamma_\ell=0.
 \label{eq:regular_requirements}
\end{equation}
Then $s=1$, $P(1,x)=1-\e^{-x}$, and
\begin{equation}
 u_r=-\frac{a(t)}2r,
 \qquad
 u_z=a(t)z+b(t),
 \label{eq:regular_meridional}
\end{equation}
\begin{equation}
 u_\theta=\frac1{2\pi r}\sum_{i=1}^N\Gamma_i
 \left[1-\e^{-\beta_i(t)r^2}\right].
 \label{eq:regular_multimode_swirl}
\end{equation}
Near the axis,
\begin{equation}
 u_\theta=\frac{r}{2\pi}\sum_i\Gamma_i\beta_i+O(r^3),
\end{equation}
so the azimuthal velocity is regular and vanishes linearly.

\subsection{Source/sink and finite-circulation classes}
When $q\ne0$, the natural domain is $r>0$. For
\begin{equation}
 s=1+\frac{q}{2\nu}>0
 \quad\Longleftrightarrow\quad q>-2\nu,
 \label{eq:spositive}
\end{equation}
the normalised incomplete-gamma form satisfies $P(s,0)=0$ and $P(s,\infty)=1$. The distributed circulation of mode $i$ inside radius $r$ is
\begin{equation}
 \Gamma_i(r,t)=\Gamma_i P(s,\beta_i r^2),
 \label{eq:distributed_circulation}
\end{equation}
with finite total distributed circulation $\Gamma_i$. For $s\le0$, the swirl is represented by the upper incomplete gamma function on admissible punctured domains. The normalised finite-circulation interpretation is restricted to $s>0$.

\subsection{Initial and boundary conditions}
The distributed KK solution is specified by $\rho$, $\nu$, constant $q$, the circulation coefficients $\Gamma_\ell,\Gamma_i$, initial radial scales $\beta_{i0}>0$, and prescribed functions $a(t)$ and $b(t)$. Defining $a_0=a(0)$ and $b_0=b(0)$, the initial field is
\begin{equation}
 u_r(r,0)=-\frac{a_0}{2}r+\frac qr,
 \qquad
 u_z(z,0)=a_0z+b_0,
\end{equation}
\begin{equation}
 u_\theta(r,0)=\frac1{2\pi r}
 \left[\Gamma_\ell+\sum_i\Gamma_iP(s,\beta_{i0}r^2)\right].
\end{equation}
For the regular branch, $u_r(0)=u_\theta(0)=0$, $|\omega_z(0)|<\infty$ and $p_r(0)=0$. At large radius,
\begin{equation}
 u_r\sim-\frac{a(t)}2r,
 \qquad
 u_z=a(t)z+b(t),
 \qquad
 u_\theta\sim\frac{\Gamma_\infty}{2\pi r}.
\end{equation}
The far-field behaviour is specified by the imposed linear axisymmetric strain and the decaying azimuthal velocity.

On a finite computational or experimental cylinder, an exact boundary-driven solution is obtained by prescribing boundary velocities or tractions from the KK field. Application to a closed cylinder with no-slip walls requires compatibility between the prescribed wall motion and the KK velocity at every boundary point.

For pointwise momentum balance on $0<t<T$, where $T$ is a prescribed terminal time, $a(t)$ and $b(t)$ are assumed continuously differentiable on every compact subinterval of $[0,T)$. Unbounded terminal behaviour is permitted as $t\to T^-$.
\section{Direct verification and intrinsic vortex structure}\label{sec:verification}

\subsection{Governing-equation residuals}
The continuity residual is defined by
\begin{equation}
 R_c=\frac1r\frac{\partial(ru_r)}{\partial r}+\frac{\partial u_z}{\partial z}.
\end{equation}
Momentum residuals $R_r$, $R_\theta$, and $R_z$ are defined by moving every term of Eqs.~\eqref{eq:radial_NS}, \eqref{eq:azimuthal_NS}, and \eqref{eq:axial_NS} to the left-hand side. For the meridional field,
\begin{equation}
 \frac1r\frac{\partial(ru_r)}{\partial r}=-a,
 \qquad
 \frac{\partial u_z}{\partial z}=a,
\end{equation}
so
\begin{equation}
 R_c=0.
\end{equation}
The radial and axial viscous operators vanish identically. Substitution of Eq.~\eqref{eq:pressure_total} therefore gives
\begin{equation}
 R_r=0,\qquad R_z=0.
\end{equation}

For one distributed angular-momentum component,
\begin{equation}
 L_i=\frac{\Gamma_i}{2\pi}P(s,x_i),
 \qquad x_i=\beta_i r^2,
\end{equation}
and the required profile derivatives are
\begin{equation}
 \frac{\partial P}{\partial x}=\frac{x^{s-1}\e^{-x}}{\Gamma(s)},
 \qquad
 \frac{\partial^2P}{\partial x^2}
 =\left(\frac{s-1}{x}-1\right)\frac{\partial P}{\partial x}.
\end{equation}
Direct substitution into Eq.~\eqref{eq:Ltransport} produces a residual proportional to
\begin{equation}
 \frac{x_i}{\beta_i}(\dot\beta_i-a\beta_i+4\nu\beta_i^2)
 +2\beta_i[q-2\nu(s-1)],
\end{equation}
which vanishes under Eqs.~\eqref{eq:beta_evolution} and \eqref{eq:sdef}. A central circulation contributes a residual proportional to $\dot\Gamma_\ell$, while a time-dependent distributed circulation contributes a term proportional to $\dot\Gamma_i$. The unforced solution therefore requires
\begin{equation}
 \dot\Gamma_\ell=0,\qquad \dot\Gamma_i=0.
 \label{eq:circulation_constant}
\end{equation}
Under these conditions,
\begin{equation}
 R_c=R_r=R_\theta=R_z=0.
 \label{eq:all_residuals_zero}
\end{equation}
The complete residual algebra is supplied in the Supplementary Material.

\subsection{Vorticity and circulation}
The vorticity vector $\bm\omega=\nabla\times\bm u$ is purely axial for the distributed field,
\begin{equation}
 \bm\omega=\omega_z\bm e_z,
\end{equation}
with
\begin{equation}
 \omega_z(r,t)=\sum_{i=1}^N
 \frac{\Gamma_i\beta_i(t)}{\pi\Gamma(s)}
 [\beta_i(t)r^2]^{s-1}\e^{-\beta_i(t)r^2}.
 \label{eq:vorticity_multimode}
\end{equation}
The line-circulation term has no classical vorticity for $r>0$; in a distributional description it corresponds to vorticity concentrated on the excluded axis.

The axial vorticity obeys
\begin{equation}
 \omega_{z,t}+
 \left(-\frac a2r+\frac qr\right)\omega_{z,r}
 =a\omega_z+\nu\left(\omega_{z,rr}+\frac1r\omega_{z,r}\right).
 \label{eq:vorticity_transport}
\end{equation}
The terms on the right separate axial stretching from viscous diffusion.

\begin{figure*}[t]
 \centering
 \includegraphics[width=0.98\textwidth]{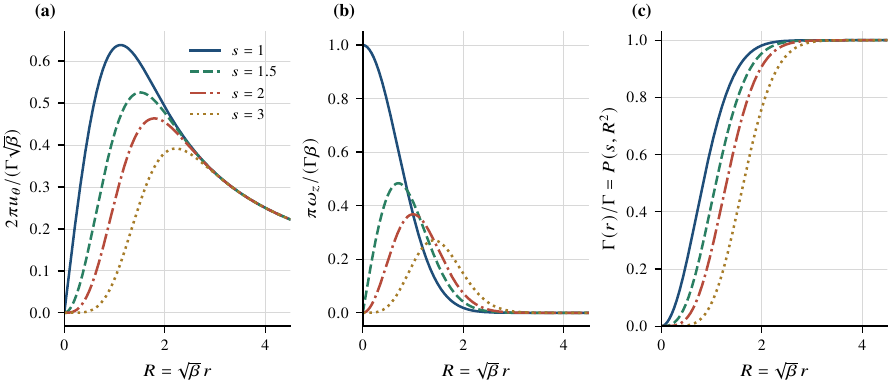}
 \caption{Single-mode finite-circulation KK profiles for representative incomplete-gamma shape parameter $s$, written in the scaled radius $R=\sqrt{\beta}\,r$. (a) Dimensionless azimuthal velocity $2\pi u_\theta/(\Gamma\sqrt{\beta})$. (b) Dimensionless axial vorticity $\pi\omega_z/(\Gamma\beta)$. (c) Fraction of total distributed circulation enclosed within radius $R$, $P(s,R^2)$. The $s=1$ curve is the regular Burgers/Oseen radial form; $s>1$ gives an annular vorticity-magnitude maximum.}
 \label{fig:profiles}
\end{figure*}

The circulation convention follows standard vortex-dynamics notation \cite{Saffman1992}. The circulation inside radius $r$ is
\begin{equation}
 \Gamma(r,t)=2\pi r u_\theta
 =\Gamma_\ell+\sum_i\Gamma_iP(s,\beta_i r^2).
 \label{eq:Gamma_r}
\end{equation}
Dimensionless velocity, vorticity and cumulative-circulation profiles for representative $s$ are shown in Fig.~\ref{fig:profiles}. Increasing $s$ shifts the distributed vorticity away from the axis and changes the swirl profile from the regular Burgers form to an annular source-bearing vortex.

\subsection{Gamma representation and radial moments}
\begin{figure*}[t]
 \centering
 \includegraphics[width=0.98\textwidth]{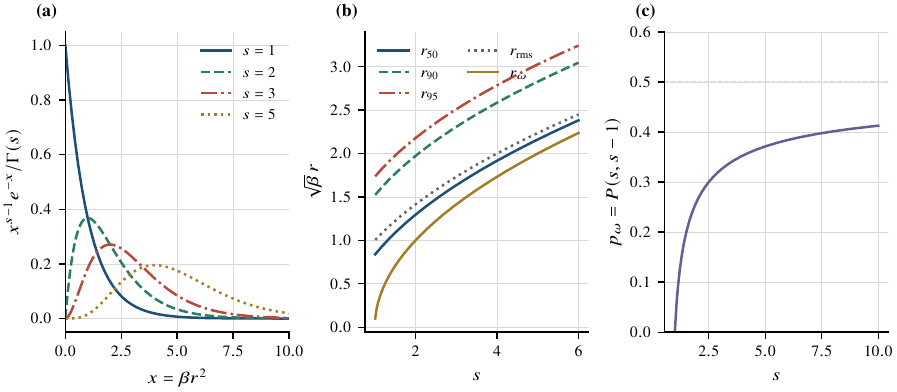}
 \caption{Gamma representation and characteristic radii of a single distributed KK mode. (a) Normalised circulation-density function in $x=\beta r^2$. (b) Scaled radii $\sqrt{\beta}\,r$ containing 50\%, 90\% and 95\% of the distributed circulation, together with the rms radius and the peak-vorticity-magnitude radius $r_\omega$. (c) Fraction $p_\omega=P(s,s-1)$ of distributed circulation enclosed by the peak-vorticity-magnitude cylinder.}
 \label{fig:gamma_structure}
\end{figure*}

For one distributed mode, $x_i=\beta_i r^2$ is introduced, and $\omega_{z,i}$ denotes its axial vorticity. Since $\dd\Gamma_i=2\pi r\omega_{z,i}\,\dd r$,
\begin{equation}
 \frac{\dd\Gamma_i}{\Gamma_i}
 =\frac{x_i^{s-1}\e^{-x_i}}{\Gamma(s)}\,\dd x_i.
 \label{eq:gamma_density}
\end{equation}
Thus the distributed circulation is represented exactly by a gamma density in scaled squared radius. For non-negative integer $n$, its even radial moment $\langle r^{2n}\rangle_i$ is
\begin{equation}
 \langle r^{2n}\rangle_i
 =\frac1{\beta_i^n}\frac{\Gamma(s+n)}{\Gamma(s)},\qquad n\ge0.
 \label{eq:moments}
\end{equation}
In particular, with $\operatorname{Var}_i(r^2)$ denoting the variance of squared radius,
\begin{equation}
 \langle r^2\rangle_i=\frac{s}{\beta_i},
 \qquad
 \operatorname{Var}_i(r^2)=\frac{s}{\beta_i^2},
\end{equation}
and
\begin{equation}
 r_{\mathrm{rms},i}=\sqrt{\frac{s}{\beta_i}}.
\end{equation}
For a prescribed circulation fraction $0<p_c<1$, let $x_{p_c}(s)$ satisfy $P(s,x_{p_c})=p_c$. The corresponding radius containing fraction $p_c$ of the distributed circulation is
\begin{equation}
 r_{p_c,i}=\sqrt{\frac{x_{p_c}(s)}{\beta_i}}.
 \label{eq:rp}
\end{equation}
These quantities are summarised in Fig.~\ref{fig:gamma_structure}. The figure also shows the peak-vorticity-magnitude circulation fraction $p_\omega$,
\begin{equation}
 p_\omega=P(s,s-1)
 \label{eq:pomega}
\end{equation}
enclosed by the peak-vorticity-magnitude surface for $s>1$.

\subsection{Peak vorticity magnitude and maximum swirl magnitude}
For one mode,
\begin{equation}
 \omega_{z,i}\propto x_i^{s-1}\e^{-x_i}.
\end{equation}
For $s>1$ and $\Gamma_i\ne0$, the maximum of $|\omega_{z,i}|$ occurs at
\begin{equation}
 x_\omega=s-1,\qquad
 r_{\omega,i}=\sqrt{\frac{s-1}{\beta_i}}
 =\sqrt{\frac{q}{2\nu\beta_i}}.
 \label{eq:romega}
\end{equation}
Its magnitude is
\begin{equation}
 |\omega_{z,i}|_{\max}
 =\frac{|\Gamma_i|\beta_i}{\pi\Gamma(s)}
 (s-1)^{s-1}\e^{-(s-1)}.
 \label{eq:omega_max}
\end{equation}
For $\Gamma_i>0$ this point is the maximum of the signed vorticity $\omega_{z,i}$; for $\Gamma_i<0$ it is the corresponding minimum.
For $s=1$, the vorticity-magnitude maximum is on the axis and $|\omega_{z,i}(0)|=|\Gamma_i|\beta_i/\pi$. The distributed vorticity is bounded at the axis only for $s\ge1$. For $0<s<1$, the distributed circulation remains finite but the vorticity diverges toward the already excluded line-sink axis.

For a single distributed component without central line circulation,
\begin{equation}
 u_{\theta,i}=\frac{\Gamma_i}{2\pi r}P(s,\beta_i r^2).
\end{equation}
For $s>1/2$ and $\Gamma_i\ne0$, the magnitude $|u_{\theta,i}|$ vanishes as $r\to0$ and $r\to\infty$ and therefore possesses a positive interior maximum. Its radial location is the positive solution $x=x_u(s)$ of
\begin{equation}
 2x^s\e^{-x}=\gamma(s,x),
 \label{eq:vmax_root}
\end{equation}
so
\begin{equation}
 r_{u,i}=\sqrt{\frac{x_u(s)}{\beta_i}}.
\end{equation}
For $\Gamma_i>0$ this is the maximum of $u_{\theta,i}$; for $\Gamma_i<0$ it is the corresponding minimum.
For $s=1$, $x_u=1.2564312086\ldots$. At $s=1/2$ the limiting axis value is finite, while for $0<s<1/2$ the azimuthal velocity magnitude diverges toward the excluded axis. In a multimode field, multiple extrema may occur; the exact extremum relation is
\begin{equation}
 2\pi r^2\omega_z(r,t)=\Gamma(r,t).
 \label{eq:multimode_vmax}
\end{equation}

\subsection{Enstrophy}
Enstrophy per unit axial length is defined as
\begin{equation}
 \Zens=\frac12\int_0^\infty\omega_z^2\,2\pi r\,\dd r.
\end{equation}
For the multimode distributed field,
\begin{equation}
 \Zens=
 \frac{\Gamma(2s-1)}{2\pi\Gamma(s)^2}
 \sum_{i=1}^N\sum_{j=1}^N
 \Gamma_i\Gamma_j
 \frac{(\beta_i\beta_j)^s}{(\beta_i+\beta_j)^{2s-1}},
 \label{eq:Zmultimode}
\end{equation}
provided $s>1/2$. The same condition is required for a finite single-mode enstrophy. For one mode,
\begin{equation}
 \Zens_i=\frac{\Gamma_i^2\beta_i\Gamma(2s-1)}{2^{2s}\pi\Gamma(s)^2}.
 \label{eq:Zsingle}
\end{equation}
Consequently,
\begin{equation}
 \dot\Zens_i=[a(t)-4\nu\beta_i(t)]\Zens_i.
 \label{eq:Zdot}
\end{equation}
The strain contribution amplifies enstrophy while viscous spreading opposes it. Constant positive strain reaches stationary balance precisely at $\beta_i=a/(4\nu)$.

\subsection{Lagrangian motion}
Let $(r_0,\theta_0,z_0)$ denote a fluid particle's initial cylindrical coordinates. Its radial position satisfies
\begin{equation}
 \dot r=-\frac{a(t)}2r+\frac qr,
\end{equation}
so
\begin{equation}
 r^2(t)=\e^{-\Aacc(t)}
 \left[r_0^2+2q\int_0^t\e^{\Aacc(\sigma)}\,\dd\sigma\right].
 \label{eq:rtraj}
\end{equation}
The axial trajectory is
\begin{equation}
 z(t)=\e^{\Aacc(t)}
 \left[z_0+\int_0^t\e^{-\Aacc(\sigma)}b(\sigma)\,\dd\sigma\right],
 \label{eq:ztraj}
\end{equation}
and the azimuthal angle follows from one exact time integral,
\begin{equation}
 \theta(t)=\theta_0+\frac1{2\pi}\int_0^t
 \frac{\Gamma_\ell+\sum_i\Gamma_iP(s,\beta_i r^2)}{r^2}\,\dd\sigma.
 \label{eq:thetatraj}
\end{equation}
\section{Exact zero-flux material circulation surface}\label{sec:material}

A material cylinder is identified in the source-bearing single-mode KK vortex for $s>1$, equivalently $q>0$, by combining the radial particle equation with the vortex-scale evolution. The derivation is valid for arbitrary admissible time-dependent strain. A constant central line circulation is permitted; its classical vorticity is zero on $r>0$.

\subsection{Scaled-radius transport}
For one distributed mode, define
\begin{equation}
 x(t)=\beta(t)r^2(t).
 \label{eq:material_x}
\end{equation}
Along a fluid trajectory,
\begin{equation}
 \frac{\mathrm D x}{\mathrm D t}
 =\dot\beta r^2+\beta\frac{\mathrm D r^2}{\mathrm D t}.
\end{equation}
Using
\begin{equation}
 \dot\beta=a\beta-4\nu\beta^2,
 \qquad
 \frac{\mathrm D r^2}{\mathrm D t}=-a r^2+2q,
\end{equation}
and $q=2\nu(s-1)$ gives
\begin{equation}
 \frac{\mathrm D x}{\mathrm D t}
 =4\nu\beta(t)\big[(s-1)-x\big].
 \label{eq:DxDt}
\end{equation}
For $s>1$, the positive scaled radius
\begin{equation}
 x_*=s-1
 \label{eq:xstar}
\end{equation}
is therefore invariant. Material points with $x<x_*$ remain inside this value and move toward it in scaled radius; material points with $x>x_*$ remain outside and also approach it in scaled radius. No trajectory crosses $x=x_*$.

An equivalent physical-space relation is obtained from $h=1/\beta$. Because
\begin{equation}
 \dot h+a h=4\nu,
\end{equation}
the radius
\begin{equation}
 r_*^2(t)=\frac{s-1}{\beta(t)}=\frac{q}{2\nu\beta(t)} 
 \label{eq:rstar}
\end{equation}
satisfies
\begin{equation}
 \frac{\mathrm d r_*^2}{\mathrm d t}=-a(t)r_*^2+2q.
 \label{eq:rstar_motion_sq}
\end{equation}
This is exactly the radial particle equation. Hence
\begin{equation}
 \frac{\mathrm d r_*}{\mathrm d t}=u_r(r_*,t).
 \label{eq:rstar_material}
\end{equation}
The same calculation yields the radial Lagrangian invariant $\mathcal I_r$,
\begin{equation}
 \mathcal I_r=
 \e^{\Aacc(t)}
 \left[r^2(t)-\frac{s-1}{\beta(t)}\right]
 =\text{constant},
 \qquad
 \Aacc(t)=\int_0^t a(\sigma)\,\dd\sigma.
 \label{eq:radial_invariant}
\end{equation}
The sign of $r^2-r_*^2$ is therefore preserved by the radial motion.

\subsection{Coincidence with the vorticity-magnitude maximum}
For a single distributed component,
\begin{equation}
 \omega_z(r,t)=
 \frac{\Gamma\beta(t)}{\pi\Gamma(s)}
 x^{s-1}\e^{-x}.
 \label{eq:material_omega}
\end{equation}
When $s>1$, differentiation gives
\begin{equation}
 \frac{\partial\omega_z}{\partial r}
 =\frac{2\Gamma\beta^{3/2}}{\pi\Gamma(s)}
 x^{s-3/2}\e^{-x}\big[(s-1)-x\big].
 \label{eq:domega_dr_material}
\end{equation}
For $\Gamma\ne0$, the unique positive-radius maximum of $|\omega_z|$ therefore occurs at $x=s-1$. Consequently,
\begin{equation}
 r_*(t)=r_\omega(t).
 \label{eq:rstar_romega}
\end{equation}
By Eq.~\eqref{eq:rstar_material}, the peak-vorticity-magnitude cylinder is transported by the radial fluid velocity for arbitrary admissible $a(t)$.

\subsection{Circulation balance on a material cylinder}
Let $R_m(t)$ denote the radius of an arbitrary material cylindrical loop and define its scaled squared radius
\begin{equation}
 x_m(t)=\beta(t)R_m^2(t).
\end{equation}
The distributed circulation $\Gamma_m(t)$ enclosed by that loop is
\begin{equation}
 \Gamma_m(t)=\Gamma P(s,x_m).
 \label{eq:material_circulation}
\end{equation}
A constant central line circulation, if present, is added to the enclosed circulation; its time derivative is zero. Using Eq.~\eqref{eq:DxDt} and
\begin{equation}
 \frac{\partial P}{\partial x}
 =\frac{x^{s-1}\e^{-x}}{\Gamma(s)},
\end{equation}
gives
\begin{equation}
 \frac{\mathrm d\Gamma_m}{\mathrm d t}
 =\frac{4\nu\beta\Gamma}{\Gamma(s)}
 x_m^{s-1}\e^{-x_m}\big[(s-1)-x_m\big].
 \label{eq:dGamma_material}
\end{equation}
The same result follows directly from the viscous angular-momentum equation. Since $L=ru_\theta$ and $\omega_z=r^{-1}L_r$,
\begin{equation}
 L_{rr}-\frac1rL_r=r\frac{\partial\omega_z}{\partial r}.
\end{equation}
Along a material loop,
\begin{equation}
 \frac{\mathrm d\Gamma_m}{\mathrm d t}
 =2\pi\nu R_m
 \frac{\partial\omega_z}{\partial r}\bigg|_{R_m}.
 \label{eq:viscous_circulation_flux}
\end{equation}
Equation~\eqref{eq:viscous_circulation_flux} is the net viscous transfer of circulation across the material cylinder.

At $R_m=r_\omega$, Eq.~\eqref{eq:domega_dr_material} gives the conserved circulation $\Gamma_*$,
\begin{equation}
 \frac{\mathrm d\Gamma_*}{\mathrm d t}=0,
 \qquad
 \Gamma_* = \Gamma P(s,s-1).
 \label{eq:zero_flux_circulation}
\end{equation}
Thus the peak-vorticity-magnitude cylinder is an exact zero-flux material circulation surface. It encloses a time-independent fraction
\begin{equation}
 \frac{\Gamma_*}{\Gamma}=P(s,s-1)
 \label{eq:circulation_fraction}
\end{equation}
of the distributed circulation. The result is independent of the imposed strain history and of the initial value of $\beta$.

For fixed non-zero circulation sign, opposite signs of the material circulation rate are obtained on either side of $x=s-1$ from Eq.~\eqref{eq:dGamma_material}. Zero viscous circulation transfer is obtained on the intervening material cylinder. The inside/outside ordering of radial trajectories is preserved by Eq.~\eqref{eq:radial_invariant}. A material partition with conserved enclosed circulation is thus established for arbitrary admissible strain histories.

The motion of the peak-vorticity-magnitude cylinder and conservation of its enclosed circulation are established by Eqs.~\eqref{eq:rstar_material} and \eqref{eq:zero_flux_circulation}, respectively. Coincidence with the stationary stagnation cylinder is recovered in the steady source--strain limit.

\begin{figure*}[t]
 \centering
 \includegraphics[width=0.98\textwidth]{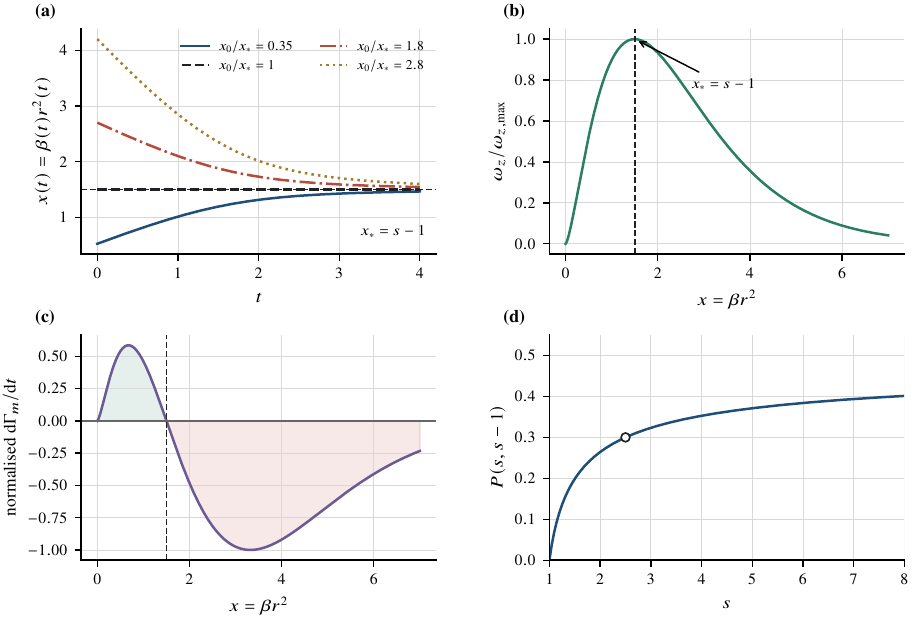}
 \caption{Exact material circulation structure of a single source-bearing KK vortex with $s>1$. (a) Scaled material radii $x(t)=\beta r^2$ approach the invariant value $x_*=s-1$ from either side; the trajectory initialized at $x_*$ remains there. (b) The distributed-vorticity magnitude is proportional to $x^{s-1}\e^{-x}$, which has its maximum at the same $x_*$. (c) The signed viscous circulation-transfer factor $x^{s-1}\e^{-x}[(s-1)-x]$ changes sign and vanishes at $x_*$. (d) The fraction of distributed circulation enclosed by the zero-flux material cylinder is $P(s,s-1)$ and depends only on $s$.}
 \label{fig:material_invariant}
\end{figure*}

The structure is illustrated in Fig.~\ref{fig:material_invariant}. The signed circulation flux changes sign at $x=s-1$. Circulation is transferred outward or inward relative to material cylinders on opposite sides, while the distinguished cylinder has zero net viscous transfer.

\begin{table*}[t]
\caption{Coupled properties of the zero-flux material circulation surface for a single distributed KK component.}\label{tab:material_invariant}
\centering
\small
\begin{tabular}{@{}p{0.23\textwidth}p{0.31\textwidth}p{0.39\textwidth}@{}}
\toprule
\raggedright Property & \raggedright Exact relation & \raggedright Fluid-mechanical implication \tabularnewline
\midrule
\raggedright Scaled-radius condition & \raggedright $x_*=s-1$ & \raggedright Positive interior surface exists for $s>1$ ($q>0$) \tabularnewline
\raggedright Material motion & \raggedright $\mathrm d r_*/\mathrm dt=u_r(r_*,t)$ & \raggedright The cylinder is transported by the radial flow \tabularnewline
\raggedright Radial invariant & \raggedright $\e^{\Aacc}[r^2-(s-1)/\beta]=\mathrm{const}$ & \raggedright Inside/outside ordering relative to $r_*$ is preserved \tabularnewline
\raggedright Vorticity-magnitude maximum & \raggedright $\partial_r\omega_z=0$ at $r=r_*$ & \raggedright The material cylinder carries peak distributed-vorticity magnitude \tabularnewline
\raggedright Viscous circulation transfer & \raggedright $\mathrm d\Gamma_m/\mathrm dt=2\pi\nu R_m\,\partial_{R_m}\omega_z$ & \raggedright Circulation changes through viscous radial transfer \tabularnewline
\raggedright Zero-flux condition & \raggedright $\mathrm d\Gamma_*/\mathrm dt=0$ & \raggedright Net viscous circulation transfer vanishes at $r_*$ \tabularnewline
\raggedright Enclosed circulation & \raggedright $\Gamma_*/\Gamma=P(s,s-1)$ & \raggedright A fixed fraction of distributed circulation is enclosed \tabularnewline
\bottomrule
\end{tabular}
\end{table*}

\subsection{Relation to circulation invariants and diffusive-transport barriers}
The conservation in Eq.~\eqref{eq:zero_flux_circulation} is a deterministic, solution-specific material-loop invariant. Generalized stochastic Kelvin formulations of viscous Navier--Stokes dynamics recover circulation conservation through stochastic Lagrangian trajectories or their averages \cite{ConstantinIyer2008,Eyink2010}. In the present solution, the ordinary material loop is transported by the deterministic KK velocity field. Its circulation is constant because the local viscous transfer term in Eq.~\eqref{eq:viscous_circulation_flux} vanishes identically on that moving loop.

In the broader variational theory, material barriers to diffusive transport are generally identified through transport functionals or variational conditions \cite{HallerKarraschKogelbauer2018}. Vortex boundaries defined by reduced diffusive vorticity leakage have been developed for two-dimensional flows \cite{KatsanoulisFarazmandSerraHaller2020}, and objective barriers to transport of dynamically active vector fields have been formulated in three dimensions \cite{HallerKatsanoulisHolznerFrohnapfelGatti2020}. The KK cylinder is identified from the common zero $x=s-1$ of the scaled material-motion, distributed-vorticity-gradient and viscous material-circulation-transfer relations. Neighbouring material cylinders have non-zero circulation-transfer rates of opposite sign for fixed non-zero circulation sign, while the intervening cylinder has exactly zero transfer. Classification of this zero-flux material circulation surface under a specified variational active-transport-barrier criterion requires evaluation of the corresponding transport functional.

\subsection{Steady limit and multimode restriction}
For constant positive strain, every positive scale approaches $\beta=a/(4\nu)$. Equation~\eqref{eq:rstar} then gives
\begin{equation}
 r_*^2=\frac{2q}{a},
\end{equation}
which is the radial stagnation radius of the steady source--strain flow. Rajamanickam and Weiss established the corresponding steady source-bearing vortex structure, including the coincidence of the radial stagnation cylinder with the vorticity-magnitude maximum \cite{RajamanickamWeiss2021}. The steady cylinder is recovered here as the stationary limit of a moving surface that remains material and carries exactly conserved circulation throughout arbitrary admissible unsteady strain histories.

For a multimode KK field, Eqs.~\eqref{eq:DxDt} and \eqref{eq:zero_flux_circulation} are satisfied separately by each distributed component with $x_i=\beta_i r^2$. The extrema and zero-flux radii of the total field are determined from the radial derivative of the summed vorticity. The componentwise derivation establishes a common material zero-flux surface for the total field in the single-mode case and in the degenerate multimode case with equal $\beta_i$, for which the profiles combine into one radial shape. For distinct scales, materiality of a total-field zero-flux surface requires a separate calculation.
\section{Structure of the KK family and classical limiting flows}\label{sec:limits}

\begin{figure*}[t]
 \centering
 \includegraphics[width=0.98\textwidth]{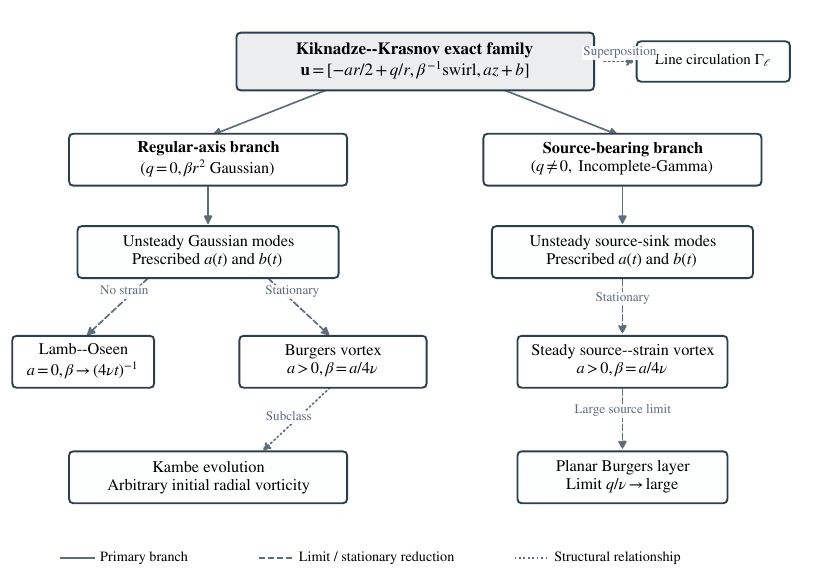}
 \caption{Solution-family map. The regular-axis KK branch ($q=0$, $\Gamma_\ell=0$) contains finite superpositions of Gaussian distributed-vorticity modes driven by a prescribed strain history. Zero strain gives Oseen spreading, while constant positive strain drives the single-scale state toward the Burgers vortex. Kambe's arbitrary-initial-vorticity solution is the broader regular strained-vorticity evolution containing the finite-Gaussian class. The source-bearing branch has incomplete-gamma structure and approaches a steady source--strain vortex under constant strain; its large-source local limit is the Burgers vortex layer.}
 \label{fig:family_map}
\end{figure*}

The principal flow classes are summarised in Fig.~\ref{fig:family_map} and Table~\ref{tab:limits}. Restricted branches, stationary states, and limiting processes are distinguished.

\begin{table*}[t]
\caption{Principal KK branches and limiting flows.}\label{tab:limits}
\centering
\begin{tabular}{@{}p{0.16\textwidth}p{0.23\textwidth}p{0.25\textwidth}p{0.28\textwidth}@{}}
\toprule
\raggedright Flow & \raggedright Restrictions / limit & \raggedright Distributed vorticity & \raggedright Physical interpretation \tabularnewline
\midrule
\raggedright Regular unsteady KK & \raggedright $q=0$, $\Gamma_\ell=0$ & \raggedright Finite Gaussian sum & \raggedright Time-dependent strained vortex \tabularnewline
\raggedright Lamb--Oseen & \raggedright Regular branch, $a=0$, $N=1$ & \raggedright Gaussian spreading & \raggedright Pure viscous diffusion \tabularnewline
\raggedright Burgers vortex & \raggedright Regular branch, $a=a_0>0$, stationary $\beta$ & \raggedright Gaussian & \raggedright Strain--diffusion equilibrium \tabularnewline
\raggedright Kambe class & \raggedright $q=0$, arbitrary regular initial vorticity & \raggedright General transformed heat-kernel solution & \raggedright Broader unsteady strained-vorticity evolution \tabularnewline
\raggedright Source-bearing KK & \raggedright $q\ne0$, $r>0$ & \raggedright Incomplete-gamma & \raggedright Strain + line-source/sink transport \tabularnewline
\raggedright Steady source--strain vortex & \raggedright $a=a_0>0$, $\beta=a_0/(4\nu)$ & \raggedright Annular for $q>0$ & \raggedright Stagnation-cylinder vortex \tabularnewline
\bottomrule
\end{tabular}
\end{table*}

\subsection{Oseen spreading under zero strain}
For $q=0$, $\Gamma_\ell=0$, $a=0$, $b=0$, and $N=1$, Equation~\eqref{eq:beta_evolution} gives
\begin{equation}
 \beta(t)=\frac1{\beta_0^{-1}+4\nu t}.
\end{equation}
Writing $t_0=(4\nu\beta_0)^{-1}$,
\begin{equation}
 u_\theta=\frac{\Gamma}{2\pi r}
 \left[1-\exp\left(-\frac{r^2}{4\nu(t+t_0)}\right)\right],
\end{equation}
\begin{equation}
 \omega_z=\frac{\Gamma}{4\pi\nu(t+t_0)}
 \exp\left(-\frac{r^2}{4\nu(t+t_0)}\right),
\end{equation}
which is the Lamb--Oseen vortex evaluated at an effective viscous age $t+t_0$ \cite{Oseen1911}.

\subsection{Burgers equilibrium under constant positive strain}
For $a(t)=a_0>0$ and the regular single-mode branch,
\begin{equation}
 \frac1{\beta(t)}=\frac{4\nu}{a_0}
 +\left(\frac1{\beta_0}-\frac{4\nu}{a_0}\right)\e^{-a_0t}.
 \label{eq:beta_constant_a}
\end{equation}
Therefore the stationary Burgers inverse squared scale $\beta_B$ is
\begin{equation}
 \beta(t)\to\beta_B=\frac{a_0}{4\nu},
\end{equation}
and
\begin{equation}
 u_{\theta,B}=\frac{\Gamma}{2\pi r}
 \left[1-\exp\left(-\frac{a_0r^2}{4\nu}\right)\right],
\end{equation}
which is the classical Burgers vortex under the adopted strain convention \cite{Burgers1948}. Its stationarity follows directly from equality of the concentration and diffusion terms in Eq.~\eqref{eq:beta_evolution}.

\subsection{Arbitrary regular initial vorticity and Kambe's solution}
For $q=0$, let $\omega(r,t)$ denote the scalar axial vorticity. Its transport equation is
\begin{equation}
 \omega_t-\frac{a(t)}2r\omega_r
 =a(t)\omega+\nu\left(\omega_{rr}+\frac1r\omega_r\right).
\end{equation}
The transformed radial coordinate $\xi$, transformed diffusion time $\tau_K$, and transformed vorticity $W$ are defined by
\begin{equation}
 \xi=r\e^{\Aacc(t)/2},
 \qquad
 \tau_K(t)=\int_0^t\e^{\Aacc(\sigma)}\,\dd\sigma,
 \qquad
 \omega(r,t)=\e^{\Aacc(t)}W(\xi,\tau_K).
\end{equation}
The equation reduces to ordinary axisymmetric diffusion,
\begin{equation}
 W_{\tau_K}=\nu\left(W_{\xi\xi}+\frac1\xi W_\xi\right).
\end{equation}
For prescribed initial vorticity $\Omega_{\mathrm{init}}(r)$,
\begin{align}
 W(\xi,\tau_K)=\int_0^\infty
 &\frac{\eta}{2\nu\tau_K}
 \exp\left[-\frac{\xi^2+\eta^2}{4\nu\tau_K}\right]\nonumber\\
 &\times I_0\left(\frac{\xi\eta}{2\nu\tau_K}\right)\Omega_{\mathrm{init}}(\eta)\,\dd\eta.
 \label{eq:Kambe_kernel}
\end{align}
Here $\eta$ is a dummy radial integration variable and $I_0$ is the modified Bessel function of the first kind of order zero.
The broader strained-vorticity solution was obtained by Kambe, with the Oseen and Burgers vortices included as special cases \cite{Kambe1984}. A finite Gaussian sum remains a finite Gaussian sum under the transformed diffusion equation and yields the regular multimode KK form. The KK source-bearing branch additionally contains $q/r$ transport and incomplete-gamma distributed vorticity.

\subsection{Steady source--strain vortex}
For constant $a_0>0$ and $q>0$, a stationary distributed mode has
\begin{equation}
 \beta=\frac{a_0}{4\nu}.
\end{equation}
The radial stagnation cylinder satisfies
\begin{equation}
 r_s^2=\frac{2q}{a_0}.
 \label{eq:rs_steady}
\end{equation}
The peak-vorticity-magnitude radius is
\begin{equation}
 r_\omega^2=\frac{s-1}{\beta}
 =\frac{q/(2\nu)}{a_0/(4\nu)}
 =\frac{2q}{a_0},
\end{equation}
so
\begin{equation}
 r_\omega=r_s.
 \label{eq:romega_rs_steady}
\end{equation}
The same relation was obtained for the steady source--strain vortex by Rajamanickam and Weiss \cite{RajamanickamWeiss2021}. For time-dependent strain, the instantaneous stagnation radius is
\begin{equation}
 r_s^2(t)=\frac{2q}{a(t)},
\end{equation}
whereas $r_\omega$ depends on the full strain history through $\beta(t)$. Their squared-radius ratio is
\begin{equation}
 \mathcal B_i(t)=\frac{r_{\omega,i}^2}{r_s^2}
 =\frac{a(t)}{4\nu\beta_i(t)}.
 \label{eq:Bbalance}
\end{equation}
The dimensionless ratio $\mathcal B_i$ measures instantaneous strain--diffusion balance. Equilibrium is marked by $\mathcal B_i=1$, and departure from unity measures the lag of the viscous vortex scale relative to the current strain.

For very large source strength the stationary cylindrical stagnation surface moves to large radius; viewed locally around that surface, curvature becomes weak and the structure approaches the Burgers vortex layer \cite{RajamanickamWeiss2021}.

\subsection{Multiscale convergence under sustained strain}
Equation~\eqref{eq:scale_separation} shows that positive accumulated strain reduces differences among inverse squared radial scales. For constant $a_0>0$,
\begin{equation}
 \beta_i(t)\to\frac{a_0}{4\nu}
\end{equation}
for every positive initial $\beta_{i0}$. Initially distinct core widths therefore approach a common strain--diffusion scale while their individual circulations remain constant.
\section{Scale dynamics and finite-time concentration}\label{sec:collapse}

The material-circulation structure of Sec.~\ref{sec:material} is controlled by the same radial scale that governs all KK distributed modes. Finite-time concentration is classified from the explicit evolution of this scale.

\subsection{Inverse scale and finite accumulated strain}
Define
\begin{equation}
 h_i(t)=\frac{1}{\beta_i(t)},
 \qquad
 \Aacc(t)=\int_0^t a(\sigma)\,\dd\sigma.
\end{equation}
Equation~\eqref{eq:beta_evolution} becomes
\begin{equation}
 \dot h_i+a(t)h_i=4\nu,
 \label{eq:h_scale}
\end{equation}
with exact solution
\begin{equation}
 h_i(t)=\e^{-\Aacc(t)}
 \left[
 h_{i0}+4\nu\int_0^t\e^{\Aacc(\sigma)}\,\dd\sigma
 \right].
 \label{eq:hcriterion}
\end{equation}
Hence, for finite $T$,
\begin{equation}
 \int_0^T |a(t)|\,\dd t<\infty
 \quad\Longrightarrow\quad
 0<\beta_i(T^-)<\infty.
 \label{eq:core_collapse_criterion}
\end{equation}
By Eq.~\eqref{eq:core_collapse_criterion}, a finite positive distributed scale is maintained up to $T$ when the imposed strain is absolutely integrable. This condition includes strain histories with an integrable terminal divergence; the distributed scale has a finite positive terminal value even when the imposed linear-strain field is singular.

The exact concentration diagnostic is
\begin{equation}
 \beta_i(t)\to\infty
 \quad\Longleftrightarrow\quad
 \e^{-\Aacc(t)}
 \left[
 \beta_{i0}^{-1}+4\nu\int_0^t\e^{\Aacc(\sigma)}\,\dd\sigma
 \right]\to0.
 \label{eq:collapse_diagnostic}
\end{equation}
Unbounded accumulated positive stretching is therefore necessary. The weighted viscous integral remains part of the condition.

\subsection{Regular strain histories}
\begin{figure*}[t]
 \centering
 \includegraphics[width=0.98\textwidth]{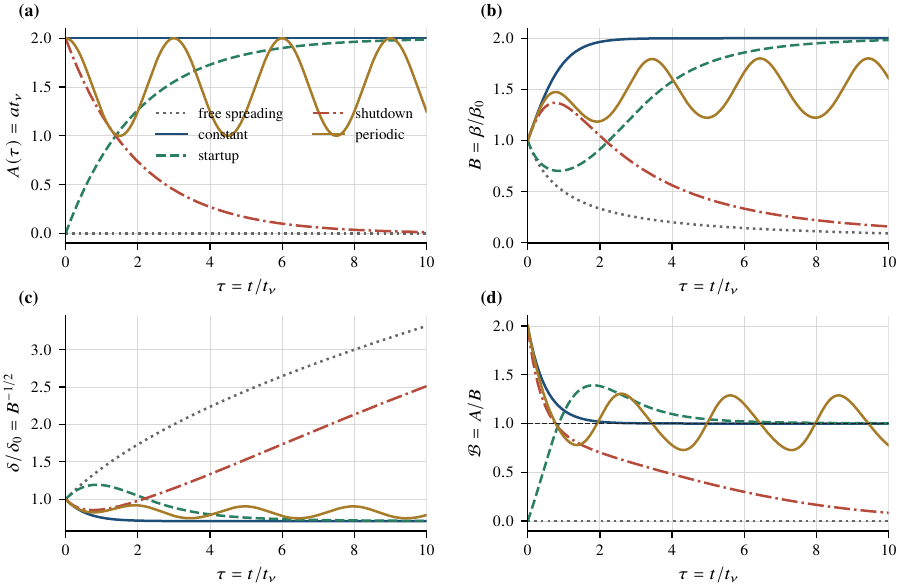}
 \caption{Dimensionless KK scale response under regular strain histories. (a) Prescribed dimensionless strain $A(\tau)$. (b) Normalised inverse squared core scale $B=\beta/\beta_0$. (c) Corresponding core-radius ratio $\delta/\delta_0=B^{-1/2}$. (d) Instantaneous strain--diffusion balance $A/B$. All histories remain finite on finite intervals.}
 \label{fig:regular_histories}
\end{figure*}

The response to bounded strain is illustrated in Fig.~\ref{fig:regular_histories}. Define the initial viscous time $t_{\nu,i}$, dimensionless time $\tau$, normalised inverse squared scale $B_i$, and dimensionless strain $A$ by
\begin{equation}
 t_{\nu,i}=\frac{1}{4\nu\beta_{i0}},
 \qquad
 \tau=\frac{t}{t_{\nu,i}},
 \qquad
 B_i=\frac{\beta_i}{\beta_{i0}},
 \qquad
 A(\tau)=a(t)t_{\nu,i},
\end{equation}
Eq.~\eqref{eq:beta_evolution} becomes
\begin{equation}
 \frac{\dd B_i}{\dd\tau}=A(\tau)B_i-B_i^2.
 \label{eq:dimensionless_beta}
\end{equation}
The characteristic core radius is defined by $\delta_i=\beta_i^{-1/2}$, with initial value $\delta_{i0}=\beta_{i0}^{-1/2}$; unsubscripted $\delta$ and $\delta_0$ denote the corresponding generic single-mode quantities. For a generic single mode, $B$ denotes the corresponding normalised scale. Zero strain gives Oseen spreading, $B=(1+\tau)^{-1}$. Constant positive strain approaches the Burgers balance. Smooth startup and shutdown remain finite for finite time. A positive-mean periodic strain produces a bounded phase-lagged response. All plotted curves are evaluated from Eq.~\eqref{eq:hcriterion} by numerical quadrature of its explicit time integral.

\subsection{Singular strain family}
Divergence of instantaneous strain and divergence of accumulated strain are compared using a family with dimensionless amplitude $\lambda>0$, terminal time $T$, and dimensionless exponent $p_s>0$,
\begin{equation}
 a(t)=\frac{\lambda}{T}\left(1-\frac{t}{T}\right)^{-p_s},
 \qquad \lambda>0,
 \quad p_s>0.
 \label{eq:singular_family}
\end{equation}
With $y=1-t/T$,
\begin{equation}
 \Aacc(t)=
 \begin{cases}
 \dfrac{\lambda}{1-p_s}\left(1-y^{1-p_s}\right), & p_s\ne1,\\[1.2ex]
 \lambda\ln(1/y), & p_s=1.
 \end{cases}
 \label{eq:singular_A}
\end{equation}
For $p_s<1$, $a(t)$ diverges but $\Aacc(T^-)$ remains finite, and the distributed scale retains a finite positive terminal value. At $p_s=1$, reciprocal-time strain produces collapse for all $\lambda>0$, with different asymptotic rates. For $\lambda>1$,
\begin{equation}
 \beta_i(t)\sim\frac{\lambda-1}{4\nu(T-t)},
 \qquad
 \delta_i(t)\sim\left[\frac{4\nu(T-t)}{\lambda-1}\right]^{1/2}.
 \label{eq:burgers_singular_scaling}
\end{equation}
This is the established singular Burgers-vortex scaling \cite{Moffatt2000,MaekawaMiuraPrange2019}. At $\lambda=1$ a logarithmic correction appears, while $0<\lambda<1$ retains a $\lambda$-dependent concentration exponent. For $p_s>1$, the leading scale follows the rapidly varying strain--diffusion balance, $\beta_i\sim a/(4\nu)$. The detailed asymptotic forms are given in the Supplementary Material.

\begin{figure*}[t]
 \centering
 \includegraphics[width=0.98\textwidth]{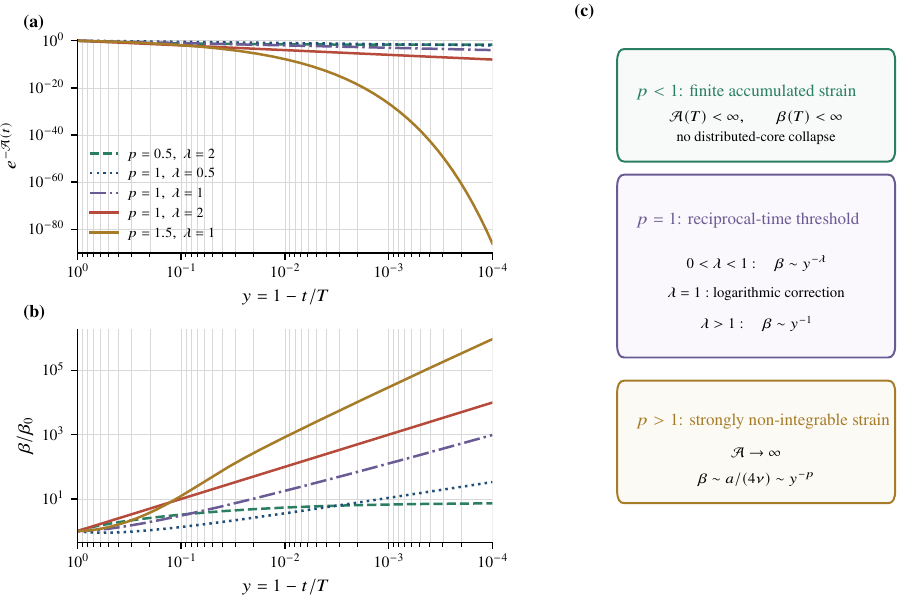}
 \caption{Finite-time KK scale response for the strain family \eqref{eq:singular_family}. (a) Attenuation factor $\exp[-\mathcal A(t)]$. (b) Growth of $\beta/\beta_0$ as the terminal time is approached. (c) Regime classification. The plot label $p$ denotes the singular-strain exponent $p_s$. For $p_s<1$, the accumulated strain remains finite despite divergent instantaneous strain. The reciprocal case $p_s=1$ separates the weaker and stronger concentration scalings, while $p_s>1$ approaches the instantaneous strain--diffusion balance. Curves are evaluated from the explicit inverse-scale relation \eqref{eq:hcriterion}.}
 \label{fig:finite_time}
\end{figure*}

\subsection{Interpretive limit}
The concentration analysis applies to the KK spatial class, whose whole-space kinetic-energy integral diverges for non-zero linear background strain. Within the KK family, a finite positive distributed scale is maintained over finite intervals under the absolute-integrability condition in Eq.~\eqref{eq:core_collapse_criterion}.

\begin{figure*}[t]
    \centering
    \includegraphics[width=0.98\textwidth]{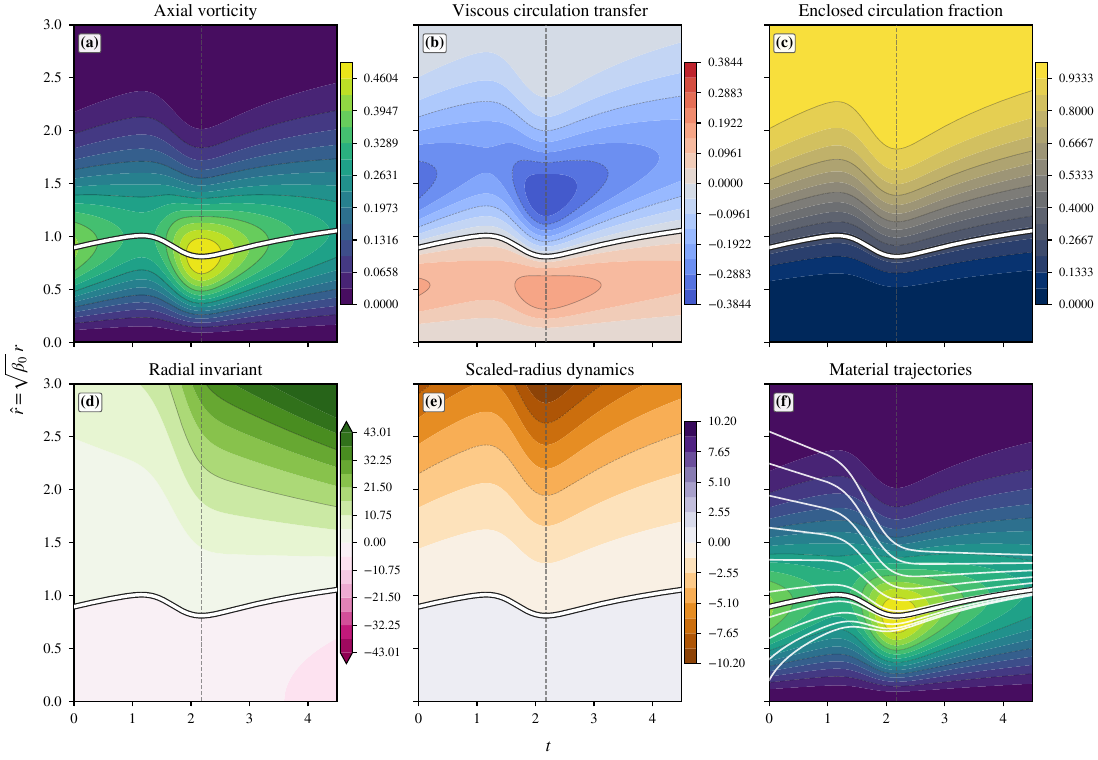}
    \caption{Exact radial--time structure of a source-bearing single-mode
    Kiknadze--Krasnov vortex under unsteady strain.
    (a) Axial vorticity; (b) viscous circulation transfer;
    (c) enclosed circulation fraction; (d) radial invariant;
    (e) scaled-radius dynamics; and (f) material trajectories over the
    vorticity field. The thick outlined curve is the exact material radius
    $\hat r_*(t)=\sqrt{\beta_0(s-1)/\beta(t)}$. Thin black curves are
    isolines, and the vertical dashed line marks the maximum of $\beta(t)$.}
    \label{fig:rt_dynamical_fields}
\end{figure*}

To visualise the unsteady transport structure in physical radius rather
than in the similarity coordinate $x=\beta r^2$, Fig.~\ref{fig:rt_dynamical_fields}
shows the exact single-mode solution in the $(t,\hat r)$ plane, where
\begin{equation}
    \hat r=\sqrt{\beta_0}\,r .
\end{equation}
The plotted fields are the normalised axial vorticity
$\pi\omega_z/(\Gamma\beta_0)$, the normalised viscous circulation-transfer
rate $\pi r\,\partial_r\omega_z/(2\Gamma\beta_0)$, the enclosed circulation
fraction $\Gamma_m/\Gamma=P(s,\beta r^2)$, the radial invariant
$\beta_0\mathcal I_r$, and the normalised scaled-radius evolution
$(4\nu\beta)^{-1}\allowbreak\,\mathrm D x/\mathrm D t$. Exact radial material
trajectories are shown separately over the vorticity field.

The same distinguished radius appears independently in all six
representations. It is the vorticity-magnitude-maximum radius
$r_*=r_\omega$, the zero of the viscous circulation-transfer rate, and
the contour enclosing the fixed fraction $P(s,s-1)$ of the distributed
circulation. It is also the zero level of the radial invariant
$\mathcal I_r$ and the fixed surface $\mathrm D x/\mathrm D t=0$.
The material trajectories in Fig.~\ref{fig:rt_dynamical_fields}(f)
remain on their respective sides of this curve, providing a direct
physical-space representation of the no-crossing property established
in Sec.~\ref{sec:material}.

\section{Discussion}\label{sec:applications}

The principal result establishes an exact compatibility between material transport and viscous circulation transport in the source-bearing single-mode branch. The distinguished moving cylinder is material, carries the distributed-vorticity-magnitude maximum, and has zero net viscous circulation transfer. Its enclosed circulation remains constant while $\beta(t)$, the vorticity amplitude, the physical core scale and neighbouring material-loop circulations evolve. The common factor $(s-1)-x$ in the material-motion, vorticity-gradient and circulation-transfer relations gives the algebraic mechanism for this compatibility. An arbitrary admissible strain history determines the physical trajectory $r_*(t)$, while the scaled distinguished value $x_*=s-1$ and the enclosed fraction $P(s,s-1)$ remain fixed.

KK streamline relations have been used to parameterize cardiovascular flow channels \cite{Beglov2025}, and the Burgers member of the regular-axis family has been used to examine tracer dispersion \cite{GoudarElsinga2018} and to prescribe circulation in a fire-whirl model \cite{ChuahKuwanaSaito2009}. In these studies, selected geometrical or transport properties were evaluated using an analytical vortex representation. This common approach motivates the use of the present reconstruction to test the dynamical consistency of prescribed vortex profiles and to quantify their evolution under imposed strain. The pressure field tests compatibility with momentum balance, the scale equation constrains core evolution, and the material-circulation invariant provides a test of circulation conservation on the specified moving surface. Their applicability to the cited settings is summarised in Table~\ref{tab:applications}.

\begin{table*}[t]
\caption{Reported applications of strained-vortex models and conditional uses of the present KK results. Proposed uses are subject to the governing assumptions and boundary conditions stated in the text.}\label{tab:applications}
\centering
\begin{tabular}{@{}p{0.18\textwidth}p{0.35\textwidth}p{0.40\textwidth}@{}}
\toprule
\raggedright Setting & \raggedright Reported basis & \raggedright Use of the present results \tabularnewline
\midrule
\raggedright Numerical verification & \raggedright Exact residual and trajectory identities derived in Secs.~\ref{sec:verification} and \ref{sec:material} & \raggedright Velocity, pressure, core evolution and material circulation can be checked with compatible boundary data \tabularnewline
\raggedright Tracer and fibre transport & \raggedright Burgers-vortex tracer calculations \cite{GoudarElsinga2018}; flexible fibres in a forced Stokes analogue \cite{Islam2026} & \raggedright Strain-dependent trajectories can be prescribed; fibre mechanics and feedback require a compatible coupled model \tabularnewline
\raggedright Cardiovascular modelling & \raggedright Trabecular morphology \cite{Bockeria2013}; KK-based channel parameterization \cite{Beglov2025}; Burgers-based vortex identification \cite{AsgharzadehBorazjani2019} & \raggedright Fitted profiles can be tested against scale and pressure relations; the annular invariant requires a source-bearing single-mode fit \tabularnewline
\raggedright Heat transfer and combustion & \raggedright Concavity-induced tornado-like jets \cite{KiknadzeGachechiladzeOleinikovAlekseev2006}; core-radius-dependent fire-whirl model \cite{ChuahKuwanaSaito2009} & \raggedright Core evolution can be prescribed in a reduced transport model with specified thermal and species balances \tabularnewline
\raggedright Atmospheric and plasma analogues & \raggedright Burgers-core dust-devil model \cite{Horton2016}; measured Burgers-type plasma vorticity \cite{Nagaoka2002} & \raggedright Profile and pressure comparisons can be made locally; thermal driving, electromagnetic forces and density variation require separate treatment \tabularnewline
\bottomrule
\end{tabular}
\end{table*}

The cardiovascular studies illustrate how geometrical evidence can be used to motivate a vortex representation. Bockeria \emph{et al.} compared trabecular orientations obtained from ventricular casts and computed tomography with KK-type streamlines \cite{Bockeria2013}. In the subsequent channel-parameterization study, anatomical surfaces of the left heart and aorta were approximated using relations associated with a quasi-stationary Burgers form \cite{Beglov2025}. An unsteady extension of such a representation can be assessed by checking that the instantaneous strain and core width satisfy Eq.~\eqref{eq:beta_evolution} throughout the motion. In particular, substitution of $\beta(t)=a(t)/(4\nu)$ into that equation leaves $\dot a/(4\nu)$, so instantaneous equilibrium is exact for constant positive strain. For general admissible strain histories, the core is determined by Eq.~\eqref{eq:beta_exact}. The pressure gradients in Eqs.~\eqref{eq:pr} and \eqref{eq:pz} then provide an additional consistency check against prescribed tractions or measured pressure differences. Extension to an anatomical domain requires the velocity to satisfy its wall motion, curvature and branching conditions.

The distinction between geometrical identification and material transport can be examined more specifically through the aneurysm calculations of Asgharzadeh and Borazjani \cite{AsgharzadehBorazjani2019}. Vortex structures in simplified geometries were delineated using a Burgers-based criterion with a vorticity magnitude equal to $\e^{-1}$ of the axial maximum. The present source-bearing single-mode solution supplies a dynamically defined comparison through the material radius $r_*^2=(s-1)/\beta$, which encloses the constant circulation $\Gamma P(s,s-1)$. These properties can be tested by following the inferred surface and evaluating its circulation over time. Application of this test requires $s>1$, constant circulation coefficients and agreement with the KK meridional velocity; the Burgers branch has $s=1$, for which $r_*=0$. Thus, use of an annular material boundary in cardiovascular data would require independent evidence for the source-bearing profile on an admissible domain. For distinct superposed scales, the total-vorticity gradient and materiality of its zero-flux surface must be evaluated separately, as established in Sec.~\ref{sec:material}.

A quantitative assessment of these requirements can be organised around simultaneous profile and evolution tests. The parameters $(\Gamma_i,\beta_i,s)$ can be estimated from radial profiles of velocity and vorticity, with $s=1+q/(2\nu)$ constrained by the meridional flow. Given an independently determined strain history, the fitted inverse scales can be tested through the residual $\dot h_i+a h_i-4\nu$ of Eq.~\eqref{eq:hlinear}. The explicit integral solution also permits comparison without differentiating noisy fitted scales. Agreement should be assessed with measurement, fitting and discretisation uncertainties propagated into the comparison. Within the fixed-profile construction, variation of the circulation coefficients requires azimuthal forcing, and variation of $q$ requires a modified swirl construction. Departures from these compatibility conditions can therefore be used to identify where the prescribed KK representation requires extension.

Before such comparisons are made with measured flows, numerical implementations of the proposed diagnostics can be checked against the exact field itself. The vanishing residuals in Eq.~\eqref{eq:all_residuals_zero} permit velocity and pressure errors to be evaluated with boundary data prescribed from the solution. A finite cylinder can be used for the regular branch and an annular domain for the source-bearing branch. The scale solution provides a reference during startup, shutdown and oscillatory strain, while Eqs.~\eqref{eq:radial_invariant} and \eqref{eq:zero_flux_circulation} provide simultaneous checks on trajectory ordering and circulation conservation for the source-bearing material cylinder. A computed departure from constant $\Gamma_*$ in this reference problem measures error in the numerical field, trajectory reconstruction or circulation evaluation. Under mesh and time-step refinement, these errors can be examined independently of uncertainty in a physical flow model.

The same trajectory checks are relevant to the tracer calculations of Goudar and Elsinga \cite{GoudarElsinga2018}, in which dispersion around a Burgers vortex was compared with dispersion around other elementary structures and in turbulence. Their Burgers-vortex results exhibited oscillatory pair dispersion and substantial departures from turbulent dispersion statistics. A corresponding use of KK is therefore a controlled test of sensitivity to strain history and radial source or sink transport. Equations~\eqref{eq:rtraj}--\eqref{eq:thetatraj} provide the particle trajectories for that test. In the source-bearing case, preservation of the sign of $r^2-r_*^2$ additionally determines whether initially separated tracer populations remain on their respective sides of the moving cylinder, provided that the tracers follow the fluid velocity exactly. The consequences for dispersion statistics would have to be evaluated for specified initial particle configurations.

For deformable objects, the particle-following assumption must be replaced by a mechanical coupling model. Bin Islam \emph{et al.} examined flexible fibres in a zero-Reynolds-number Burgers-like flow generated by regularised Stokes singularities, with fibre forces included in the fluid problem \cite{Islam2026}. Fibre deformation depended on the elastoviscous number and the ratio of vortex-core diameter to fibre length. This dependence motivates examining how a prescribed time-dependent core modifies fibre response. The KK scale equation could specify such a core history in a compatible fluid--structure formulation, with the fluid regime and fibre feedback treated explicitly. Under a negligible-feedback approximation, the KK velocity could be supplied as the carrier field; fibre trajectories and deformation would still be obtained from the fibre equations. The material-circulation invariant applies to loops advected by the carrier fluid, so its use for extended objects requires a separate assessment of their motion relative to those loops.

Transport of a dissolved or thermal quantity introduces a related dependence on its own governing equation. For a passive scalar $\phi$ with constant diffusivity $D_\phi>0$,
\begin{equation}
 \phi_t+\bm u\cdot\nabla\phi=D_\phi\nabla^2\phi,
\end{equation}
the velocity can be prescribed from KK when scalar feedback is negligible. Scalar exchange across a material surface is then determined by the normal diffusive flux $-D_\phi\nabla\phi\cdot\bm n$. On the cylinder $r=r_*$, zero scalar flux therefore requires $\partial_r\phi=0$; the vorticity condition $\partial_r\omega_z=0$ constrains circulation transfer alone. This separation of transport balances is necessary when the material cylinder is used as a reference for mixing or heat-transfer calculations.

The dependence of thermal transport on vortex structure is illustrated by the fire-whirl model of Chuah, Kuwana and Saito \cite{ChuahKuwanaSaito2009}. Species and energy balances were coupled to a Burgers circulation field, and the predicted heat feedback and flame geometry depended on the vortex-core radius. The present scale solution supplies a candidate time-dependent core history for extending that type of reduced model. Evaluation of the extension would require the unsteady species and energy equations and renewed comparison with thermal measurements, with momentum coupling included where density variation or buoyancy modifies the prescribed flow. Tornado-like jets induced by concave surface reliefs have also been analysed by Kiknadze \emph{et al.} \cite{KiknadzeGachechiladzeOleinikovAlekseev2006}. For this setting, local velocity profiles and core evolution could be compared with KK before a scalar-transport model is introduced; wall geometry and thermal boundary conditions would remain part of the application-specific problem.

The need to determine the driving mechanism independently is also evident in atmospheric and plasma applications. In the dust-devil model of Horton \emph{et al.}, a Burgers core was embedded in a thermally driven flow whose development was investigated numerically \cite{Horton2016}. Nagaoka \emph{et al.} measured a Burgers-type vorticity distribution around a density cavity in a rotating magnetized plasma \cite{Nagaoka2002}. These results motivate local comparisons of core profiles across different physical systems. The KK pressure and scale relations provide conditional hydrodynamic references for such comparisons when constant density, viscosity and the prescribed meridional flow are adequate approximations. Thermal generation of the atmospheric circulation and electromagnetic forcing and density variation in the plasma must be included in their respective governing models. A measured Burgers-type profile supports a comparison with the regular branch; use of the source-bearing invariant would additionally require its meridional and material-motion conditions to be satisfied.

Across these proposed uses, the exact scale dynamics constrain how vortex concentration may be interpreted. If the imposed strain is absolutely integrable over a finite interval, Eq.~\eqref{eq:core_collapse_criterion} preserves a finite positive distributed scale. This provides a consistency test for apparent core collapse in a KK-based computation or fitted evolution. Singular imposed strain, prescribed line singularities and distributed-core concentration can be distinguished through $a(t)$, $(q,\Gamma_\ell)$ and $\beta_i(t)$, respectively. The non-zero linear background strain gives infinite whole-space kinetic energy, so the concentration analysis is restricted to that spatial class. Within the admissible source-bearing branch, the circulation invariant supplies an additional constraint throughout the scale evolution. A proposed application can therefore be assessed by checking its boundary and forcing assumptions, its fitted scale dynamics and, where $s>1$, the motion and circulation of the candidate material surface.
\section{Conclusions}\label{sec:conclusions}

The Kiknadze--Krasnov axisymmetric velocity field has been derived from the incompressible mass and momentum balances and expressed in circulation-based notation. Axial strain, radial line-source or line-sink transport, central circulation and distributed incomplete-gamma swirl are included in the velocity field. The associated pressure field and compatibility conditions have been obtained, and all Navier--Stokes residuals have been verified directly. This reconstruction provides the analytical setting for a deterministic material-loop invariant in the source-bearing single-mode family. Viscous circulation transfer vanishes persistently on the distinguished loop during unsteady evolution. The principal result concerns a single source-bearing distributed component with $s>1$. Its scaled squared radius $x=\beta r^2$ satisfies
\begin{equation}
 \frac{\mathrm D x}{\mathrm D t}=4\nu\beta[(s-1)-x].
\end{equation}
The cylinder $x=s-1$ is therefore material for arbitrary admissible strain. In physical coordinates,
\begin{equation}
 r_*^2=\frac{s-1}{\beta(t)},
 \qquad
 \frac{\mathrm d r_*}{\mathrm d t}=u_r(r_*,t),
\end{equation}
and
\begin{equation}
 \e^{\Aacc(t)}\left[r^2-r_*^2\right]=\mathrm{constant}
\end{equation}
along every radial trajectory. The same cylinder is the location of maximum distributed-vorticity magnitude. The circulation enclosed by a material cylinder obeys
\begin{equation}
 \frac{\mathrm d\Gamma_m}{\mathrm d t}
 =2\pi\nu R_m\frac{\partial\omega_z}{\partial r}\bigg|_{R_m}.
\end{equation}
At $R_m=r_*$ the vorticity gradient vanishes, so
\begin{equation}
 \frac{\mathrm d\Gamma_*}{\mathrm d t}=0,
 \qquad
 \Gamma_*=\Gamma P(s,s-1).
\end{equation}
The peak-vorticity-magnitude cylinder is therefore identified as an exact zero-flux material circulation surface by the material-motion and viscous circulation identities. A fixed fraction of the distributed circulation is enclosed for every imposed strain history compatible with the KK solution.

Selective material-circulation conservation is established within the exact KK family. Neighbouring material cylinders generally exchange circulation viscously, as Eq.~\eqref{eq:dGamma_material} shows. The distinguished deterministic material loop retains constant circulation throughout the admissible unsteady evolution through the persistent compatibility of its material motion with zero viscous circulation transfer.

In the stationary state under constant positive strain, the steady source--strain result is recovered with $\beta=a/(4\nu)$ and $r_*^2=2q/a$, and the vorticity-magnitude maximum is located on the radial stagnation cylinder. In a general multimode KK field, the material-circulation identities are satisfied componentwise. A common total-field material zero-flux surface is established when the radial scales are equal. For distinct scales, the total-vorticity extrema and zero-flux radii must be determined from the summed vorticity, and materiality must be checked separately.

The reconstruction also fixes the parameter restrictions of the finite-circulation branch. For nonzero single-mode circulation, an interior maximum of $|u_\theta|$ requires $s>1/2$. Finite single-mode enstrophy requires $s>1/2$. Distributed vorticity is bounded at the axis only for $s\ge1$, and the non-trivial annular material peak-vorticity-magnitude surface requires $s>1$.

With $h_i=1/\beta_i$, the squared radial scale is obtained explicitly as
\begin{equation}
 h_i(t)=\e^{-\Aacc(t)}
 \left[h_{i0}+4\nu\int_0^t\e^{\Aacc(\sigma)}\,\dd\sigma\right].
\end{equation}
An initially finite positive distributed scale is preserved over finite intervals when the imposed strain is absolutely integrable. The established Burgers-vortex concentration scaling is recovered under reciprocal-time singular strain in the appropriate limit. The concentration result is restricted to the KK family, whose non-zero linear background strain gives infinite whole-space kinetic energy.

The persistent deterministic material-loop invariant extends the known steady coincidence of radial stagnation and the vorticity-magnitude maximum to admissible unsteady evolution within the exact source-bearing single-mode KK family and the equal-scale reductions stated above. Stochastic Kelvin formulations concern circulation transported through stochastic Lagrangian trajectories or their averages. Stability under departures from the KK family remains an open question. Variational active-transport-barrier classification requires evaluation of the corresponding transport functional. Numerical circulation transport can be assessed against the conserved quantity $\Gamma P(s,s-1)$, and computed core-scale evolution can be compared with the explicit solution for $h_i(t)$. Reduced-order applications are restricted by the prescribed axisymmetry, constant fluid properties and compatible boundary data.

\section*{Supplementary material}
A separate Supplementary Material PDF contains the notation concordance, complete Navier--Stokes residual algebra, expanded derivation of the material-circulation identities, parameter-domain audit, multimode qualification, finite-time asymptotics, and computational verification details. A Google Colab notebook is supplied separately and reproduces the symbolic checks and vector figures. Formal verification of the mathematical results in Lean 4 and Mathlib is provided in the accompanying versioned repository \cite{Saqr2026KKLean}.

\section*{Data and code availability}
All numerical values shown in the figures are evaluations of the analytical relations reported in the manuscript. The supplementary \href{https://colab.research.google.com/drive/1SD6s61jgEvCe2-LxCUwIHd0sVnC7h1wu?usp=sharing}{Google Colab notebook} regenerates the symbolic verification, numerical quadratures, and all vector figures from these analytical inputs. The Lean 4 and Mathlib formal-verification companion, including machine-checked certificates for the principal material-circulation result and the supporting analytical results, is available in the versioned GitHub repository \cite{Saqr2026KKLean}.

\section*{Competing interests}
The author declares no competing interests.

\makeatletter
\if@filesw
  \immediate\write\@auxout{\string\citation{AIPAuthorControl}}
\fi
\makeatother
\renewcommand{\bibsection}{\section*{References}}
\bibliography{references}

@CONTROL{AIPAuthorControl,
  author = {48}
}

@article{BellamyKnights1970,
  author  = {Bellamy-Knights, P. G.},
  title   = {An unsteady two-cell vortex solution of the {Navier--Stokes} equations},
  journal = {Journal of Fluid Mechanics},
  volume  = {41},
  number  = {3},
  pages   = {673--687},
  year    = {1970},
  doi     = {10.1017/S0022112070000836}
}

@article{Craik2009,
  author  = {Craik, Alex D. D.},
  title   = {Exact vortex solutions of the {Navier--Stokes} equations with axisymmetric strain and suction or injection},
  journal = {Journal of Fluid Mechanics},
  volume  = {626},
  pages   = {291--306},
  year    = {2009},
  doi     = {10.1017/S0022112009005849}
}

@article{ConstantinIyer2008,
  author  = {Constantin, Peter and Iyer, Gautam},
  title   = {A stochastic {Lagrangian} representation of the three-dimensional incompressible {Navier--Stokes} equations},
  journal = {Communications on Pure and Applied Mathematics},
  volume  = {61},
  number  = {3},
  pages   = {330--345},
  year    = {2008},
  doi     = {10.1002/cpa.20192}
}

@article{Eyink2010,
  author  = {Eyink, Gregory L.},
  title   = {Stochastic least-action principle for the incompressible {Navier--Stokes} equation},
  journal = {Physica D: Nonlinear Phenomena},
  volume  = {239},
  number  = {14},
  pages   = {1236--1240},
  year    = {2010},
  doi     = {10.1016/j.physd.2008.11.011}
}

@article{Oseen1911,
  author  = {Oseen, C. W.},
  title   = {{\"U}ber Wirbelbewegung in einer reibenden Fl{\"u}ssigkeit},
  journal = {Arkiv f{\"o}r Matematik, Astronomi och Fysik},
  volume  = {7},
  number  = {14},
  pages   = {1--13},
  year    = {1911}
}

@incollection{Burgers1948,
  author    = {Burgers, J. M.},
  title     = {A mathematical model illustrating the theory of turbulence},
  booktitle = {Advances in Applied Mechanics},
  volume    = {1},
  pages     = {171--199},
  publisher = {Elsevier},
  year      = {1948},
  doi       = {10.1016/S0065-2156(08)70100-5},
  note      = {{DOI}: \url{https://doi.org/10.1016/S0065-2156(08)70100-5}}
}

@article{Kambe1984,
  author  = {Kambe, Tsutomu},
  title   = {Axisymmetric vortex solution of Navier--Stokes equation},
  journal = {Journal of the Physical Society of Japan},
  volume  = {53},
  number  = {1},
  pages   = {13--15},
  year    = {1984},
  doi     = {10.1143/JPSJ.53.13},
  note    = {{DOI}: \url{https://doi.org/10.1143/JPSJ.53.13}}
}

@article{KiknadzeKrasnov1986,
  author  = {Kiknadze, G. I. and Krasnov, Yu. K.},
  title   = {Evolution of a spout-like flow of a viscous fluid},
  journal = {Soviet Physics Doklady},
  volume  = {31},
  number  = {10},
  pages   = {799--801},
  year    = {1986}
}

@article{Ohkitani2004,
  author  = {Ohkitani, Koji},
  title   = {A survey on a class of exact solutions of the Navier--Stokes equations and a model for turbulence},
  journal = {Publications of the Research Institute for Mathematical Sciences},
  volume  = {40},
  number  = {4},
  pages   = {1267--1290},
  year    = {2004},
  doi     = {10.2977/PRIMS/1145475447},
  note    = {{DOI}: \url{https://doi.org/10.2977/PRIMS/1145475447}}
}

@article{Moffatt2000,
  author  = {Moffatt, H. K.},
  title   = {The interaction of skewed vortex pairs: a model for blow-up of the Navier--Stokes equations},
  journal = {Journal of Fluid Mechanics},
  volume  = {409},
  pages   = {51--68},
  year    = {2000},
  doi     = {10.1017/S002211209900782X},
  note    = {{DOI}: \url{https://doi.org/10.1017/S002211209900782X}}
}

@article{MaekawaMiuraPrange2019,
  author  = {Maekawa, Yasunori and Miura, Hideyuki and Prange, Christophe},
  title   = {On stability of blow-up solutions of the Burgers vortex type for the Navier--Stokes equations with a linear strain},
  journal = {Journal of Mathematical Fluid Mechanics},
  volume  = {21},
  number  = {4},
  pages   = {46},
  year    = {2019},
  doi     = {10.1007/s00021-019-0450-5},
  note    = {{DOI}: \url{https://doi.org/10.1007/s00021-019-0450-5}}
}

@article{RajamanickamWeiss2021,
  author  = {Rajamanickam, Prabakaran and Weiss, Adam D.},
  title   = {Steady axisymmetric vortices in radial stagnation flows},
  journal = {The Quarterly Journal of Mechanics and Applied Mathematics},
  volume  = {74},
  number  = {3},
  pages   = {367--378},
  year    = {2021},
  doi     = {10.1093/qjmam/hbab009},
  note    = {{DOI}: \url{https://doi.org/10.1093/qjmam/hbab009}}
}

@inproceedings{Kiknadze2009,
  author    = {Kiknadze, G. I. and Gachechiladze, I. A. and Gorodkov, A. Yu.},
  title     = {Self-organization of tornado-like jets in flows of gases and liquids and the technologies utilizing this phenomenon},
  booktitle = {Proceedings of the ASME 2009 Heat Transfer Summer Conference},
  pages     = {547--560},
  year      = {2009},
  doi       = {10.1115/HT2009-88644},
  note      = {{DOI}: \url{https://doi.org/10.1115/HT2009-88644}}
}

@article{Beglov2025,
  author  = {Beglov, Ivan and Talygin, Eugene and Zharkov, Yaroslav and Gorodkov, Alexandre},
  title   = {A new approach to the parameterization of streamlined surfaces of the left heart and aorta, taking into account the twisted structure of blood flow},
  journal = {Applied Sciences},
  volume  = {15},
  number  = {10},
  pages   = {5291},
  year    = {2025},
  doi     = {10.3390/app15105291},
  note    = {{DOI}: \url{https://doi.org/10.3390/app15105291}}
}

@article{GoudarElsinga2018,
  author  = {Goudar, Manu V. and Elsinga, Gerrit E.},
  title   = {Tracer particle dispersion around elementary flow patterns},
  journal = {Journal of Fluid Mechanics},
  volume  = {843},
  pages   = {872--897},
  year    = {2018},
  doi     = {10.1017/jfm.2018.146},
  note    = {{DOI}: \url{https://doi.org/10.1017/jfm.2018.146}}
}

@article{Islam2026,
  author  = {Bin Islam, Rubaiyat and Morshed, Adnan and Cortez, Ricardo and Fauci, Lisa},
  title   = {Dancing fibres in a microscale Burgers-like vortex},
  journal = {Journal of Fluid Mechanics},
  volume  = {1032},
  pages   = {A7},
  year    = {2026},
  doi     = {10.1017/jfm.2026.11342},
  note    = {{DOI}: \url{https://doi.org/10.1017/jfm.2026.11342}}
}

@article{AsgharzadehBorazjani2019,
  author  = {Asgharzadeh, Hafez and Borazjani, Iman},
  title   = {A non-dimensional parameter for classification of the flow in intracranial aneurysms. I. Simplified geometries},
  journal = {Physics of Fluids},
  volume  = {31},
  number  = {3},
  pages   = {031904},
  year    = {2019},
  doi     = {10.1063/1.5033942},
  note    = {{DOI}: \url{https://doi.org/10.1063/1.5033942}}
}

@article{ChuahKuwanaSaito2009,
  author  = {Chuah, Keng Hoo and Kuwana, Kazunori and Saito, Kozo},
  title   = {Modeling a fire whirl generated over a 5-cm-diameter methanol pool fire},
  journal = {Combustion and Flame},
  volume  = {156},
  number  = {9},
  pages   = {1828--1833},
  year    = {2009},
  doi     = {10.1016/j.combustflame.2009.06.010},
  note    = {{DOI}: \url{https://doi.org/10.1016/j.combustflame.2009.06.010}}
}

@article{Horton2016,
  author  = {Horton, W. and Miura, H. and Onishchenko, O. and Couedel, L. and Arnas, C. and Escarguel, A. and Benkadda, S. and Fedun, V.},
  title   = {Dust devil dynamics},
  journal = {Journal of Geophysical Research: Atmospheres},
  volume  = {121},
  number  = {12},
  pages   = {7197--7214},
  year    = {2016},
  doi     = {10.1002/2016JD024832},
  note    = {{DOI}: \url{https://doi.org/10.1002/2016JD024832}}
}

@article{Nagaoka2002,
  author  = {Nagaoka, Kenichi and Okamoto, Atsushi and Yoshimura, Shinji and Kono, Mitsuo and Tanaka, Masayoshi Y.},
  title   = {Spontaneous formation of a plasma hole in a rotating magnetized plasma: a giant Burgers vortex in a compressible fluid},
  journal = {Physical Review Letters},
  volume  = {89},
  number  = {7},
  pages   = {075001},
  year    = {2002},
  doi     = {10.1103/PhysRevLett.89.075001},
  note    = {{DOI}: \url{https://doi.org/10.1103/PhysRevLett.89.075001}}
}

@article{HallerKarraschKogelbauer2018,
  author  = {Haller, George and Karrasch, Daniel and Kogelbauer, Florian},
  title   = {Material barriers to diffusive and stochastic transport},
  journal = {Proceedings of the National Academy of Sciences of the United States of America},
  volume  = {115},
  number  = {37},
  pages   = {9074--9079},
  year    = {2018},
  doi     = {10.1073/pnas.1720177115},
  note    = {{DOI}: \url{https://doi.org/10.1073/pnas.1720177115}}
}

@article{KatsanoulisFarazmandSerraHaller2020,
  author  = {Katsanoulis, Stergios and Farazmand, Mohammad and Serra, Mattia and Haller, George},
  title   = {Vortex boundaries as barriers to diffusive vorticity transport in two-dimensional flows},
  journal = {Physical Review Fluids},
  volume  = {5},
  number  = {2},
  pages   = {024701},
  year    = {2020},
  doi     = {10.1103/PhysRevFluids.5.024701},
  note    = {{DOI}: \url{https://doi.org/10.1103/PhysRevFluids.5.024701}}
}

@article{HallerKatsanoulisHolznerFrohnapfelGatti2020,
  author  = {Haller, George and Katsanoulis, Stergios and Holzner, Markus and Frohnapfel, Bettina and Gatti, Davide},
  title   = {Objective barriers to the transport of dynamically active vector fields},
  journal = {Journal of Fluid Mechanics},
  volume  = {905},
  pages   = {A17},
  year    = {2020},
  doi     = {10.1017/jfm.2020.737},
  note    = {{DOI}: \url{https://doi.org/10.1017/jfm.2020.737}}
}

@book{Saffman1992,
  author    = {Saffman, P. G.},
  title     = {Vortex Dynamics},
  publisher = {Cambridge University Press},
  address   = {Cambridge},
  year      = {1993},
  doi       = {10.1017/CBO9780511624063},
  note      = {{DOI}: \url{https://doi.org/10.1017/CBO9780511624063}}
}

@article{Bockeria2013,
  author  = {Bockeria, Leo A. and Kiknadze, Gennady I. and Gachechiladze, Ivan A. and Gabidullina, Rosalia F. and Makarenko, Vladimir N. and Gorodkov, Alexander Y.},
  title   = {Analysis of structure of intraventricular blood flow based on studies of architectonics of trabecular layer in left ventricle},
  journal = {Cardiometry},
  number  = {3},
  pages   = {5--30},
  year    = {2013},
  doi     = {10.12710/cardiometry.2013.3.530},
  note    = {{DOI}: \url{https://doi.org/10.12710/cardiometry.2013.3.530}}
}

@article{KiknadzeGachechiladzeOleinikovAlekseev2006,
  author  = {Kiknadze, G. I. and Gachechiladze, I. A. and Oleinikov, V. G. and Alekseev, V. V.},
  title   = {Mechanisms of the self-organization of tornado-like jets flowing past three-dimensional concave reliefs},
  journal = {Heat Transfer Research},
  volume  = {37},
  number  = {6},
  pages   = {467--494},
  year    = {2006},
  doi     = {10.1615/HeatTransRes.v37.i6.10},
  note    = {{DOI}: \url{https://doi.org/10.1615/HeatTransRes.v37.i6.10}}
}

@misc{Saqr2026KKLean,
  author       = {Saqr, Khalid M.},
  title        = {Formal Verification Companion to {A Zero-Flux Material Cylinder in the Unsteady Kiknadze--Krasnov Vortex}},
  year         = {2026},
  howpublished = {GitHub repository, version 1.0.0},
  url          = {https://github.com/KNOWDYN/Kiknadze-Krasnov},
  note         = {{Lean 4} and {Mathlib} formal verification}
}
\clearpage
\onecolumngrid
\begingroup
\def\KKSupplementInMain{}
\ifdefined\KKSupplementInMain
  \let\KKSupplementFinish\relax
\else
\documentclass[aip,pof,preprint,amsmath,amssymb]{revtex4-2}
\usepackage[T1]{fontenc}
\usepackage[utf8]{inputenc}
\usepackage{bm,mathtools}
\usepackage{booktabs}

\makeatletter
\auto@bib@empty
\let\write@bibliographystyle\relax
\makeatother

\newcommand{\dd}{\mathrm{d}}
\newcommand{\e}{\mathrm{e}}
\newcommand{\Aacc}{\mathcal{A}}
\newcommand{\Zens}{\mathcal{Z}}
\newcommand{\KKSupplementFinish}{\end{document}}
\begin{document}
\fi

\renewcommand{\arraystretch}{1.12}
\setlength{\tabcolsep}{6pt}
\allowdisplaybreaks[2]
\renewcommand{\theequation}{S\arabic{equation}}
\renewcommand{\thetable}{S\arabic{table}}
\providecommand{\texorpdfstring}[2]{#1}

\ifdefined\KKSupplementInMain
  \setcounter{equation}{0}
  \setcounter{table}{0}
  \setcounter{section}{0}
  \renewcommand{\theHequation}{supp.\arabic{equation}}
  \renewcommand{\theHtable}{supp.\arabic{table}}
  \renewcommand{\theHsection}{supp.\arabic{section}}
\fi

\begin{center}
{\Large\bfseries Supplementary Material}\par
\vspace{4pt}
{\large\bfseries A Zero-Flux Material Cylinder in the Unsteady Kiknadze--Krasnov Vortex}\par
\vspace{8pt}
{\normalsize Khalid M. Saqr}\par
\vspace{3pt}
{\small Mechanical Engineering Department, College of Engineering and Technology,\par
Arab Academy for Science, Technology and Maritime Transport, Alexandria 1029, Egypt\par
\texttt{k.saqr@aast.edu}\quad ORCID: 0000-0002-3058-2705}
\end{center}
\vspace{6pt}
\hrule
\vspace{10pt}

The derivations below establish the notation equivalences, governing-equation residuals, material-surface identities, parameter domains, multimode qualifications, finite-time asymptotics, and reproducibility checks for the Kiknadze--Krasnov solution.

\section{Notation concordance}

\subsection{Equivalent parameterisations}

Positive $a(t)$ denotes axial stretching. In the coefficient-based KK parameterisation, a single-mode velocity field is
\begin{align}
 u_r&=C_0(t)r+\frac{C_1}{r},
 &u_z&=-2C_0(t)z+C_2(t),\\
 u_\theta&=\frac{1}{r}\left[
 A_1+A_2\Gamma\!\left(1+\frac{C_1}{2\nu},\alpha(t)r^2\right)
 \right].
\end{align}
The equivalent circulation-based parameters are
\begin{align}
 a(t)&=-2C_0(t), & q&=C_1, & b(t)&=C_2(t),\\
 s&=1+\frac{C_1}{2\nu}, & \beta(t)&=\alpha(t).&&
\end{align}
Writing $2\pi r u_\theta=K_0+K_1\Gamma(s,x)$ and using $\Gamma(s,x)=\Gamma(s)[1-P(s,x)]$ gives
\begin{align}
 \Gamma_1&=-K_1\Gamma(s)=-2\pi A_2\Gamma(s),\\
 \Gamma_\ell&=K_0+K_1\Gamma(s)
 =2\pi\bigl[A_1+A_2\Gamma(s)\bigr].
\end{align}
The far-field circulation is $\Gamma_\infty=K_0=2\pi A_1$.

\clearpage
\onecolumngrid
\small
\setlength{\abovedisplayskip}{8pt plus 2pt minus 2pt}
\setlength{\belowdisplayskip}{8pt plus 2pt minus 2pt}
\setlength{\abovedisplayshortskip}{4pt plus 2pt}
\setlength{\belowdisplayshortskip}{6pt plus 2pt minus 2pt}
\setlength{\jot}{3pt}

\section{Governing-equation verification}

For
\begin{equation}
 u_r=-\frac{a}{2}r+\frac{q}{r},
 \qquad
 u_z=az+b,
\end{equation}
the continuity residual is
\begin{equation}
 R_c=\frac{1}{r}(ru_r)_r+u_{z,z}=-a+a=0.
\end{equation}
The radial viscous operator is
\begin{align}
 u_{r,rr}+\frac{1}{r}u_{r,r}-\frac{u_r}{r^2}
 &=\frac{2q}{r^3}
 +\frac{1}{r}\left(-\frac{a}{2}-\frac{q}{r^2}\right)
 -\frac{1}{r^2}\left(-\frac{a}{2}r+\frac{q}{r}\right)\\
 &=0,
\end{align}
and the axial viscous operator vanishes because $u_z$ is linear in $z$ and independent of $r$.

The non-swirl radial acceleration is
\begin{align}
 u_{r,t}+u_ru_{r,r}
 &=-\frac{\dot a}{2}r+
 \left(-\frac{a}{2}r+\frac{q}{r}\right)
 \left(-\frac{a}{2}-\frac{q}{r^2}\right)\\
 &=\left(-\frac{\dot a}{2}+\frac{a^2}{4}\right)r-\frac{q^2}{r^3}.
\end{align}
The axial acceleration is
\begin{equation}
 u_{z,t}+u_zu_{z,z}=(\dot a+a^2)z+(\dot b+ab).
\end{equation}
These expressions are balanced exactly by the corresponding pressure gradients, with the centrifugal term $u_\theta^2/r$ retained in the radial balance.

For one distributed angular-momentum component,
\begin{equation}
 L=\frac{\Gamma}{2\pi}P(s,x),
 \qquad
 x=\beta r^2,
\end{equation}
\begin{equation}
 P_x=\frac{x^{s-1}\e^{-x}}{\Gamma(s)},
 \qquad
 P_{xx}=\left(\frac{s-1}{x}-1\right)P_x.
\end{equation}
Substitution into
\begin{equation}
 L_t+u_rL_r=\nu\left(L_{rr}-\frac{1}{r}L_r\right)
\end{equation}
leaves the factor
\begin{equation}
 \frac{x}{\beta}(\dot\beta-a\beta+4\nu\beta^2)
 +2\beta\bigl[q-2\nu(s-1)\bigr].
\end{equation}
The residual therefore vanishes when
\begin{equation}
 \dot\beta-a\beta+4\nu\beta^2=0,
 \qquad
 q=2\nu(s-1).
\end{equation}
A time-dependent circulation amplitude would leave an additional non-zero term; hence $\Gamma$ and $\Gamma_\ell$ are constant in the unforced solution.

\section{Material-surface derivation}

\subsection{Scaled-radius evolution}

For a radial material trajectory,
\begin{equation}
 \frac{\mathrm D r^2}{\mathrm D t}=-a(t)r^2+2q.
\end{equation}
With $x=\beta r^2$,
\begin{align}
 \frac{\mathrm D x}{\mathrm D t}
 &=\dot\beta r^2+\beta\frac{\mathrm D r^2}{\mathrm D t}\\
 &=(a\beta-4\nu\beta^2)r^2+\beta(-ar^2+2q)\\
 &=2q\beta-4\nu\beta x\\
 &=4\nu\beta\bigl[(s-1)-x\bigr].
\end{align}
The positive fixed scaled radius is $x_*=s-1$ for $s>1$.

\subsection{Physical-space invariant}

Let $h=1/\beta$. The scale equation gives
\begin{equation}
 \dot h=-a h+4\nu.
\end{equation}
Since $q=2\nu(s-1)$,
\begin{equation}
 \frac{\dd}{\dd t}\left(\frac{s-1}{\beta}\right)
 =-a\frac{s-1}{\beta}+2q.
\end{equation}
Subtracting this relation from the particle equation yields
\begin{equation}
 \frac{\dd}{\dd t}\left[r^2-\frac{s-1}{\beta}\right]
 =-a(t)\left[r^2-\frac{s-1}{\beta}\right].
\end{equation}
With $\Aacc(t)=\int_0^t a(\sigma)\,\dd\sigma$,
\begin{equation}
 \e^{\Aacc(t)}\left[r^2-\frac{s-1}{\beta(t)}\right]
 =r_0^2-\frac{s-1}{\beta_0},
\end{equation}
where $\Aacc(0)=0$. The sign of the bracket is preserved, and the surface $r^2=(s-1)/\beta$ is material.

\subsection{Vorticity-magnitude maximum and circulation transfer}

For one distributed mode,
\begin{equation}
 \omega_z=\frac{\Gamma\beta}{\pi\Gamma(s)}x^{s-1}\e^{-x}.
\end{equation}
Using $\partial_r x=2\beta r$,
\begin{align}
 \partial_r\omega_z
 &=\frac{2\Gamma\beta^2r}{\pi\Gamma(s)}
 x^{s-2}\e^{-x}\bigl[(s-1)-x\bigr]\\
 &=\frac{2\Gamma\beta^{3/2}}{\pi\Gamma(s)}
 x^{s-3/2}\e^{-x}\bigl[(s-1)-x\bigr].
\end{align}
For $s>1$, the maximum of the distributed-vorticity magnitude is therefore located at $x=s-1$.

For a material cylinder of radius $R_m(t)$,
\begin{equation}
 \Gamma_m=\Gamma P(s,x_m),
 \qquad
 x_m=\beta R_m^2.
\end{equation}
Hence
\begin{align}
 \frac{\dd\Gamma_m}{\dd t}
 &=\Gamma P_x\frac{\mathrm D x_m}{\mathrm D t}\\
 &=\frac{4\nu\beta\Gamma}{\Gamma(s)}
 x_m^{s-1}\e^{-x_m}\bigl[(s-1)-x_m\bigr].
\end{align}
The same expression follows from angular-momentum diffusion. Since $\omega_z=L_r/r$,
\begin{equation}
 L_{rr}-\frac{1}{r}L_r=r\,\partial_r\omega_z.
\end{equation}
The material derivative of $2\pi L$ therefore gives
\begin{equation}
 \frac{\dd\Gamma_m}{\dd t}=2\pi\nu R_m\,\partial_{R_m}\omega_z.
\end{equation}
At $R_m=r_\omega$,
\begin{equation}
 \frac{\dd\Gamma_*}{\dd t}=0,
 \qquad
 \Gamma_*=\Gamma P(s,s-1).
\end{equation}
A constant line circulation adds a constant to $\Gamma_m$ and leaves this derivative unchanged.

\section{Parameter domains}

The relevant restrictions are distinct and should not be interchanged.
\begin{center}
\small
\textbf{Table S1. Parameter restrictions for the finite-circulation single-mode branch.}\par\medskip
\begin{tabular}{@{}p{0.14\textwidth}p{0.34\textwidth}p{0.43\textwidth}@{}}
\toprule
\raggedright Condition & \raggedright Quantity & \raggedright Consequence \tabularnewline
\midrule
\raggedright $s>0$ & \raggedright $P(s,x)$ finite-circulation representation & \raggedright $P(s,0)=0$, $P(s,\infty)=1$ \tabularnewline
\raggedright $s>1/2$ & \raggedright $u_\theta$ near the axis & \raggedright Interior maximum of $|u_\theta|$; finite single-mode enstrophy \tabularnewline
\raggedright $s\ge1$ & \raggedright $\omega_z$ near the axis & \raggedright Distributed vorticity remains bounded \tabularnewline
\raggedright $s=1$ & \raggedright Regular $q=0$ branch & \raggedright Gaussian vorticity with an on-axis magnitude maximum \tabularnewline
\raggedright $s>1$ & \raggedright Source-bearing branch & \raggedright Annular maximum of $|\omega_z|$ and a non-trivial material zero-flux cylinder \tabularnewline
\raggedright $0<s<1$ & \raggedright Line-sink branch on $r>0$ & \raggedright Finite distributed circulation but divergent vorticity toward the excluded axis \tabularnewline
\bottomrule
\end{tabular}
\end{center}

For $s>1/2$, the maximum-$|u_\theta|$ condition follows from
\begin{equation}
 u_\theta\propto\frac{P(s,x)}{\sqrt{x}},
\end{equation}
whose small-$x$ behaviour is $x^{s-1/2}$. The interior extremum satisfies
\begin{equation}
 2x^s\e^{-x}=\gamma(s,x).
\end{equation}
The single-mode enstrophy is
\begin{equation}
 \Zens=\frac{\Gamma^2\beta\Gamma(2s-1)}{2^{2s}\pi\Gamma(s)^2},
 \qquad
 s>\frac{1}{2}.
\end{equation}

\section{Multimode qualification}

The angular-momentum equation is linear after the meridional field is fixed. A finite sum of distributed components is therefore exact:
\begin{equation}
 \omega_z=\sum_i\omega_{z,i},
 \qquad
 \Gamma(r,t)=\Gamma_\ell+\sum_i\Gamma_iP(s,\beta_i r^2).
\end{equation}
Each component separately satisfies
\begin{equation}
 \frac{\mathrm D x_i}{\mathrm D t}
 =4\nu\beta_i\bigl[(s-1)-x_i\bigr],
\end{equation}
and its own component circulation has zero viscous transfer on $x_i=s-1$. In general,
\begin{equation}
 \partial_r\omega_z=\sum_i\partial_r\omega_{z,i}
\end{equation}
does not vanish at an individual $r_{*,i}$. The peak magnitude of the total vorticity and the zero-flux surface of the total physical circulation must therefore be determined from the summed field. The single-mode result applies directly to the total field. It also applies in the degenerate equal-scale case, $\beta_i=\beta$, because the distributed components then share one radial profile and may be combined into a single circulation amplitude.

\section{Finite-time scale asymptotics}

The exact inverse scale is
\begin{equation}
 h(t)=\e^{-\Aacc(t)}
 \left[h_0+4\nu\int_0^t\e^{\Aacc(\sigma)}\,\dd\sigma\right].
\end{equation}
For
\begin{equation}
 a(t)=\frac{\lambda}{T}\left(1-\frac{t}{T}\right)^{-p_s},
 \qquad
 y=1-\frac{t}{T},
\end{equation}
\begin{equation}
 \Aacc(t)=
 \begin{cases}
 \dfrac{\lambda}{1-p_s}\bigl(1-y^{1-p_s}\bigr), & p_s\ne1,\\[1ex]
 \lambda\ln(1/y), & p_s=1.
 \end{cases}
\end{equation}

\subsection{\texorpdfstring{$0<p_s<1$}{0 < p\_s < 1}}

The accumulated strain remains finite:
\begin{equation}
 \Aacc(T^-)=\frac{\lambda}{1-p_s}<\infty.
\end{equation}
Therefore $0<\beta(T^-)<\infty$ even though $a(t)\to\infty$.

\subsection{\texorpdfstring{$p_s=1$}{p\_s = 1}}

For $\lambda\ne1$,
\begin{equation}
 h(t)=y^\lambda
 \left[h_0+4\nu T\frac{1-y^{1-\lambda}}{1-\lambda}\right].
\end{equation}
If $0<\lambda<1$,
\begin{equation}
 \beta(t)\sim
 \frac{y^{-\lambda}}{h_0+4\nu T/(1-\lambda)}.
\end{equation}
If $\lambda=1$,
\begin{equation}
 h(t)=y\bigl[h_0+4\nu T\ln(1/y)\bigr],
\end{equation}
so
\begin{equation}
 \beta(t)\sim
 \frac{1}{4\nu(T-t)\ln\!\left[T/(T-t)\right]}.
\end{equation}
If $\lambda>1$,
\begin{equation}
 \beta(t)\sim\frac{\lambda-1}{4\nu(T-t)},
\end{equation}
which yields the Burgers-type core scaling $\delta\propto(T-t)^{1/2}$.

\subsection{\texorpdfstring{$p_s>1$}{p\_s > 1}}

The accumulated strain diverges faster than logarithmically. Laplace-endpoint balance of the explicit integral gives
\begin{equation}
 \beta(t)\sim\frac{a(t)}{4\nu}
 =\frac{\lambda}{4\nu T}y^{-p_s}.
\end{equation}
The leading radial scale is therefore the instantaneous strain--diffusion scale and loses memory of $\beta_0$.

\section{Reproducibility checks}

The companion computational notebook verifies the governing-equation residuals, incomplete-gamma identities, enstrophy integral, and material-circulation relations directly from the analytical expressions. No spatial Navier--Stokes discretisation or ODE integration is required; regular and singular strain histories are obtained from the explicit inverse-scale relation by numerical quadrature. The zero-flux checks evaluate
\begin{equation}
 \frac{\mathrm D x}{\mathrm D t}=4\nu\beta\bigl[(s-1)-x\bigr],
 \quad \omega_z\propto x^{s-1}\e^{-x},
 \quad \frac{\dd\Gamma_m}{\dd t}\propto
 x^{s-1}\e^{-x}\bigl[(s-1)-x\bigr].
\end{equation}
Vector PDF outputs and the corresponding numerical arrays are generated from these relations for independent checking.
\clearpage
\onecolumngrid
\section{Nomenclature}

The following glossary covers the notation used in the article and this supplement. Local integration variables and equation-specific derivatives are identified at their first occurrence. An em dash denotes entries without a separately specified unit.

\begingroup
\linespread{1}\fontsize{10}{12}\selectfont
\setlength{\tabcolsep}{4pt}
\setlength{\parskip}{0pt}
\renewcommand{\arraystretch}{1.08}
\newcommand{\nomenclatureentry}[3]{
  \par\noindent
  \begin{tabular}[t]{@{}p{\dimexpr0.22\linewidth-2\tabcolsep\relax}p{\dimexpr0.62\linewidth-2\tabcolsep\relax}p{0.16\linewidth}@{}}
  \raggedright #1 & \raggedright #2 & \raggedright #3 \tabularnewline
  \end{tabular}
  \par\vspace{2pt}
}
\newcommand{\nomenclatureheading}[1]{
  \par\addvspace{10pt}
  \noindent\textbf{#1}\par\nopagebreak[4]
  \noindent\begin{tabular}{@{}p{\dimexpr0.22\linewidth-2\tabcolsep\relax}p{\dimexpr0.62\linewidth-2\tabcolsep\relax}p{0.16\linewidth}@{}}
  \toprule
  \raggedright\textbf{Symbol} & \raggedright\textbf{Definition} & \raggedright\textbf{SI unit} \tabularnewline
  \midrule
  \end{tabular}\par\nopagebreak[4]
}
\nomenclatureheading{Roman symbols and operators}
\nomenclatureentry{$A_1,A_2$}{KK coefficients associated with the line and distributed circulation amplitudes.}{m$^2$\,s$^{-1}$}
\nomenclatureentry{$a(t),a_0$}{Prescribed axial strain rate and its initial or constant value; $a>0$ denotes axial stretching.}{s$^{-1}$}
\nomenclatureentry{$A(\tau)$}{Dimensionless strain, $A=a t_{\nu,i}$.}{---}
\nomenclatureentry{$\Aacc(t)$}{Accumulated strain, $\Aacc(t)=\int_0^t a(\sigma)\,\dd\sigma$.}{---}
\nomenclatureentry{$b(t),b_0$}{Uniform axial-velocity component and its initial value.}{m\,s$^{-1}$}
\nomenclatureentry{$B_i,B$}{Modewise and generic normalised inverse squared radial scales, $B_i=\beta_i/\beta_{i0}$ and $B=\beta/\beta_0$.}{---}
\nomenclatureentry{$\mathcal B_i$}{Instantaneous strain--diffusion balance ratio, $\mathcal B_i=a/(4\nu\beta_i)$.}{---}
\nomenclatureentry{$C_0(t)$}{KK radial-strain coefficient, $C_0=-a/2$.}{s$^{-1}$}
\nomenclatureentry{$C_1,C_2(t)$}{KK coefficients corresponding to the line-source strength $q$ and axial translation $b(t)$.}{m$^2$\,s$^{-1}$ and m\,s$^{-1}$}
\nomenclatureentry{$\bm D$}{Rate-of-strain tensor, $\bm D=[\nabla\bm u+(\nabla\bm u)^{\mathsf T}]/2$.}{s$^{-1}$}
\nomenclatureentry{$D_\phi$}{Constant molecular diffusivity of a passive scalar.}{m$^2$\,s$^{-1}$}
\nomenclatureentry{$\bm e_r,\bm e_\theta,\bm e_z$}{Cylindrical unit vectors in the radial, azimuthal and axial directions.}{---}
\nomenclatureentry{$F(x),G(x)$}{Angular-momentum profile and its derivative, $G=F'$.}{---}
\nomenclatureentry{$h(t),h_i(t)$}{Generic and modewise squared scale variables, $h=\beta^{-1}$ and $h_i=\beta_i^{-1}$.}{m$^2$}
\nomenclatureentry{$h_0,h_{i0}$}{Generic and modewise initial squared scales.}{m$^2$}
\nomenclatureentry{$I_0$}{Modified Bessel function of the first kind of order zero.}{---}
\nomenclatureentry{$\bm I$}{Identity tensor.}{---}
\nomenclatureentry{$\mathcal I_r$}{Radial Lagrangian invariant, $\mathcal I_r=\e^{\Aacc}[r^2-(s-1)/\beta]$.}{m$^2$}
\nomenclatureentry{$K,K_0,K_1$}{Constants of integration in the angular-momentum similarity solution.}{m$^2$\,s$^{-1}$}
\nomenclatureentry{$L(r,t)$}{Specific axial angular momentum, $L=ru_\theta$.}{m$^2$\,s$^{-1}$}
\nomenclatureentry{$N$}{Number of distributed vortex modes.}{---}
\nomenclatureentry{$\bm n$}{Outward unit normal to a control surface.}{---}
\nomenclatureentry{$p,p_0(t)$}{Mechanical pressure and arbitrary time-dependent pressure reference.}{Pa}
\nomenclatureentry{$p_c$}{Prescribed fraction of distributed circulation used to define $x_{p_c}$ and $r_{p_c,i}$.}{---}
\nomenclatureentry{$p_s$}{Dimensionless exponent in the singular-strain family.}{---}
\nomenclatureentry{$p_\omega$}{Fraction of distributed circulation enclosed by the peak-vorticity-magnitude cylinder, $p_\omega=P(s,s-1)$.}{---}
\nomenclatureentry{$P(s,x)$}{Regularised lower incomplete gamma function, $P(s,x)=\gamma(s,x)/\Gamma(s)$.}{---}
\nomenclatureentry{$q$}{Constant radial line-source or line-sink strength; $q>0$ denotes an outward source.}{m$^2$\,s$^{-1}$}
\nomenclatureentry{$Q_r$}{Radial volume flux per unit axial length, $Q_r=2\pi q$.}{m$^2$\,s$^{-1}$}
\nomenclatureentry{$r,\theta,z$}{Cylindrical coordinates.}{m, rad, m}
\nomenclatureentry{$R$}{Scaled plotting radius, $R=\sqrt{\beta}\,r$.}{---}
\nomenclatureentry{$R_i$}{Radius of a finite inner cylinder used in a punctured domain.}{m}
\nomenclatureentry{$R_m(t)$}{Radius of an arbitrary material cylindrical loop.}{m}
\nomenclatureentry{$R_c,R_r,R_\theta,R_z$}{Continuity and cylindrical momentum residuals.}{---}
\nomenclatureentry{$r_*$}{Radius of the zero-flux material cylinder, $r_*^2=(s-1)/\beta$.}{m}
\nomenclatureentry{$r_{*,i}$}{Componentwise zero-flux material-cylinder radius for mode $i$.}{m}
\nomenclatureentry{$r_\omega,r_{\omega,i}$}{Generic and modewise radii of maximum distributed-vorticity magnitude.}{m}
\nomenclatureentry{$r_s$}{Radial stagnation radius in the source--strain flow.}{m}
\nomenclatureentry{$r_u,r_{u,i}$}{Generic and modewise radii of the interior maximum of $|u_\theta|$.}{m}
\nomenclatureentry{$r_{p_c,i}$}{Radius enclosing fraction $p_c$ of the distributed circulation of mode $i$.}{m}
\nomenclatureentry{$r_{\mathrm{ref}}$}{Lower radial reference used in the swirl-pressure integral.}{m}
\nomenclatureentry{$r_{\mathrm{rms},i}$}{Root-mean-square radius of mode $i$.}{m}
\nomenclatureentry{$s$}{Incomplete-gamma shape parameter, $s=1+q/(2\nu)$.}{---}
\nomenclatureentry{$t,t_0$}{Time and effective initial viscous age of the Lamb--Oseen vortex.}{s}
\nomenclatureentry{$t_{\nu,i}$}{Initial viscous scale time, $t_{\nu,i}=(4\nu\beta_{i0})^{-1}$.}{s}
\nomenclatureentry{$T$}{Terminal time in the singular-strain family.}{s}
\nomenclatureentry{$\bm u$}{Velocity vector, $\bm u=u_r\bm e_r+u_\theta\bm e_\theta+u_z\bm e_z$.}{m\,s$^{-1}$}
\nomenclatureentry{$u_r,u_\theta,u_z$}{Radial, azimuthal and axial velocity components.}{m\,s$^{-1}$}
\nomenclatureentry{$V,\partial V$}{Fixed control volume and its boundary.}{---}
\nomenclatureentry{$W(\xi,\tau_K)$}{Transformed vorticity in the regular-axis diffusion representation.}{s$^{-1}$}
\nomenclatureentry{$\mathcal W(r,t)$}{Circulation function, $\mathcal W=2\pi r u_\theta$.}{m$^2$\,s$^{-1}$}
\nomenclatureentry{$x,x_i,x_m$}{Scaled squared radii, $x=\beta r^2$, $x_i=\beta_i r^2$, and $x_m=\beta R_m^2$.}{---}
\nomenclatureentry{$x_*,x_\omega$}{Invariant material-cylinder and peak-vorticity-magnitude locations; both equal $s-1$.}{---}
\nomenclatureentry{$x_u,x_{p_c}$}{Scaled location of maximum $|u_\theta|$ and scaled radius enclosing circulation fraction $p_c$.}{---}
\nomenclatureentry{$X,Y$}{Cartesian coordinates in the transverse plane, $X=r\cos\theta$ and $Y=r\sin\theta$.}{m}
\nomenclatureentry{$\hat r,\hat r_*$}{Normalised physical radius, $\hat r=\sqrt{\beta_0}r$, and its material-cylinder value.}{---}
\nomenclatureentry{$y$}{Time-to-terminal variable, $y=1-t/T$.}{---}
\nomenclatureheading{Greek symbols and special functions}
\nomenclatureentry{$\alpha(t)$}{KK inverse squared radial scale, equivalent to $\beta(t)$.}{m$^{-2}$}
\nomenclatureentry{$\beta(t),\beta_i(t)$}{Generic and modewise inverse squared radial scales.}{m$^{-2}$}
\nomenclatureentry{$\beta_0,\beta_{i0}$}{Generic and modewise initial inverse squared radial scales.}{m$^{-2}$}
\nomenclatureentry{$\beta_B$}{Stationary Burgers inverse squared radial scale, $\beta_B=a_0/(4\nu)$.}{m$^{-2}$}
\nomenclatureentry{$\chi$}{Dimensionless source parameter, $\chi=q/(2\nu)=s-1$.}{---}
\nomenclatureentry{$\delta_i,\delta$}{Modewise and generic characteristic core radii, equal to $\beta_i^{-1/2}$ and $\beta^{-1/2}$.}{m}
\nomenclatureentry{$\delta_{i0},\delta_0$}{Modewise and generic initial core radii.}{m}
\nomenclatureentry{$\Gamma(s)$}{Complete gamma function; context distinguishes it from circulation amplitudes.}{---}
\nomenclatureentry{$\gamma(s,x),\Gamma(s,x)$}{Lower and upper incomplete gamma functions.}{---}
\nomenclatureentry{$\Gamma$}{Generic total circulation of a single distributed mode.}{m$^2$\,s$^{-1}$}
\nomenclatureentry{$\Gamma_\ell$}{Central line circulation.}{m$^2$\,s$^{-1}$}
\nomenclatureentry{$\Gamma_i$}{Total distributed circulation of mode $i$; $\Gamma_1$ is its single-mode form.}{m$^2$\,s$^{-1}$}
\nomenclatureentry{$\Gamma_i(r,t)$}{Distributed circulation of mode $i$ enclosed within radius $r$.}{m$^2$\,s$^{-1}$}
\nomenclatureentry{$\Gamma_\infty$}{Total far-field circulation, $\Gamma_\infty=\Gamma_\ell+\sum_i\Gamma_i$.}{m$^2$\,s$^{-1}$}
\nomenclatureentry{$\Gamma(r,t)$}{Total circulation enclosed within radius $r$.}{m$^2$\,s$^{-1}$}
\nomenclatureentry{$\Gamma_m,\Gamma_*$}{Circulation enclosed by an arbitrary material loop and by the zero-flux material cylinder.}{m$^2$\,s$^{-1}$}
\nomenclatureentry{$\lambda$}{Dimensionless amplitude of the singular-strain family.}{---}
\nomenclatureentry{$\mu$}{Dynamic viscosity.}{Pa\,s}
\nomenclatureentry{$\nu$}{Kinematic viscosity.}{m$^2$\,s$^{-1}$}
\nomenclatureentry{$\xi,\eta$}{Transformed and integration radii in Kambe's diffusion representation.}{m}
\nomenclatureentry{$\Pi(r,t)$}{Swirl-induced pressure potential.}{m$^2$\,s$^{-2}$}
\nomenclatureentry{$\rho$}{Constant fluid density.}{kg\,m$^{-3}$}
\nomenclatureentry{$\sigma,\bm\sigma$}{Dummy time variable in temporal integrals (s) and Cauchy stress tensor (Pa), respectively.}{---}
\nomenclatureentry{$\tau_K,\tau$}{Kambe transformed diffusion time (s) and dimensionless viscous time $t/t_{\nu,i}$.}{---}
\nomenclatureentry{$\phi$}{Passive scalar.}{---}
\nomenclatureentry{$\bm\omega,\omega,\omega_z,\omega_{z,i}$}{Vorticity vector, scalar axial vorticity in the regular-axis representation, total axial vorticity, and axial vorticity of mode $i$.}{s$^{-1}$}
\nomenclatureentry{$\Omega_{\mathrm{init}}(r)$}{Prescribed initial vorticity in the regular-axis Kambe representation.}{s$^{-1}$}
\nomenclatureentry{$\Omega_0,\Omega_q$}{Regular-axis and punctured spatial domains.}{---}
\nomenclatureentry{$\Zens,\Zens_i$}{Total and modewise enstrophy per unit axial length.}{m$^2$\,s$^{-2}$}
\nomenclatureheading{Indices, derivatives and statistical notation}
\nomenclatureentry{$i,j$}{Distributed-mode indices.}{---}
\nomenclatureentry{$n$}{Non-negative integer moment order.}{---}
\nomenclatureentry{$r_0,z_0,\theta_0$}{Initial radial, axial and azimuthal particle coordinates.}{m, m, rad}
\nomenclatureentry{$\mathrm D/\mathrm Dt$}{Material derivative following a fluid trajectory.}{---}
\nomenclatureentry{$\mathrm d/\mathrm dt$}{Ordinary total derivative with respect to time.}{---}
\nomenclatureentry{$\partial_t,\partial_r,\partial_z$}{Partial derivatives with respect to $t$, $r$, and $z$; corresponding field subscripts have the same meaning.}{---}
\nomenclatureentry{$\dot{(\,)}$}{Time derivative of the enclosed quantity.}{---}
\nomenclatureentry{$\langle r^{2n}\rangle_i$}{Even radial moment of the distributed circulation of mode $i$.}{m$^{2n}$}
\nomenclatureentry{$\operatorname{Var}_i(r^2)$}{Variance of $r^2$ for mode $i$.}{m$^4$}
\endgroup

\KKSupplementFinish

\endgroup
\end{document}